\documentclass[final,3p,times,11pt]{elsarticle}

\usepackage[USenglish]{babel}
\usepackage{amsmath,amssymb,amsthm, mathrsfs}
\usepackage{etoolbox}
\usepackage{mathtools}
\usepackage{graphicx}
\usepackage{stmaryrd}
\usepackage[dvipsnames]{xcolor}
\usepackage{cancel}
\usepackage{ulem}
\usepackage{tabularx}
\usepackage{comment}
\usepackage[linesnumbered,ruled,vlined]{algorithm2e}
\definecolor{Myblue}{rgb}{.2 0.4 1}

\usepackage{hyperref}
\hypersetup{
    colorlinks = true,       
}

\usepackage{marginnote}
\usepackage{multirow}

\newtheorem{theorem}{Theorem}
\newtheorem{lemma}{Lemma}
\AfterEndEnvironment{theorem}{\noindent\ignorespaces}
\newtheorem{remark}{Remark}
\AfterEndEnvironment{remark}{\noindent\ignorespaces}

\journal{}
\makeatletter
\def\ps@pprintTitle{%
 \let\@oddhead\@empty
 \let\@evenhead\@empty
 \def\@oddfoot{}%
 \let\@evenfoot\@oddfoot}
\makeatother

\begin{document}
\begin{frontmatter}


\title{Surrogate-based multilevel Monte Carlo methods for uncertainty quantification in the Grad-Shafranov free boundary problem }



\author[umdcs]{Howard C.\ Elman}
\ead{helman@umd.edu}
\address[umdcs]{Department of Computer Science and Institute for Advanced Computer
Studies, University of Maryland, College Park.}
\author[umdm]{Jiaxing Liang}
\ead{jl508@rice.edu}
\address[umdm]{Department of Computational Applied Mathematics \& Operations Research, Rice University.}
\author[UA]{Tonatiuh S\'anchez-Vizuet}
\ead{tonatiuh@arizona.edu}
\address[UA]{Department of Mathematics, The University of Arizona. \\[2ex] \textbf{In memory of Prof. Howard C. Elman, esteemed friend, mentor, and outstanding computational scientist}}


\begin{abstract}
 We explore a hybrid technique to quantify the variability in the numerical solutions to a free boundary problem associated with magnetic equilibrium in axisymmetric fusion reactors amidst parameter uncertainties. The method aims at reducing computational costs by integrating a surrogate model into a multilevel Monte Carlo method. The resulting surrogate-enhanced multilevel Monte Carlo methods reduce the cost of simulation by factors as large as $10^4$ compared to standard Monte Carlo simulations involving direct numerical solutions of the associated Grad-Shafranov partial differential equation. Accuracy assessments also show that surrogate-based sampling closely aligns with the results of direct computation, confirming its effectiveness in capturing the behavior of plasma boundary and geometric descriptors.
\end{abstract}

\begin{keyword}
Multilevel Monte Carlo Finite-Element \sep Sparse Grid Stochastic Collocation \sep Uncertainty Quantification \sep Grad-Shafranov Free Boundary Problem.
\MSC[2020] 65Z05 \sep 65C05
\sep  62P35 \sep 35R35 \sep 35R60.
\end{keyword}
\end{frontmatter}


\section{Introduction}
The Grad-Shafranov free boundary problem describes the static equilibrium state of a plasma in an axially symmetric magnetic confinement reactor. The mathematical model involves several parameters related to physical quantities that are either measured experimentally, inferred statistically, or subject to variability. The inherent uncertainties in these parameters pose challenges to the accurate prediction of the plasma behavior and the confinement properties of the external magnetic field. Addressing these challenges requires efficient computational methods that can efficiently handle the uncertainties and provide reliable statistical analyses of the plasma's response to varying conditions. 

The Monte Carlo (MC) method is typically employed for this purpose due to its agnosticism with respect to the dimensionality of the parameter space, which, however, comes at the cost of a notoriously slow convergence rate. This method requires gathering a large number of samples from which to infer statistical properties of the quantities of interest. In our context, obtaining a sample requires a ``direct'' numerical solution of a discretized nonlinear partial differential equation; a process that can quickly become computationally expensive. To improve Monte Carlo sampling efficiency in the context of magnetic plasma confinement, our initial efforts \cite{ElLiSa:2022} involved the use of the sparse grid stochastic collocation method to build a surrogate function that would alleviate the need for multiple direct solutions. Said surrogate is constructed by solving the discrete system for specific values of the parameters and using this information to build a high-order interpolant of the solution operator which is thereafter used for sampling. Expanding on this work, in \cite{ElLiSa:2023} we then explored a different sampling strategy aimed at enhancing the efficiency of MC -- the multilevel Monte Carlo (MLMC) method \cite{BaScZo:2011,Gi:2008}. During the sampling step, rather than collecting a large number of solutions obtained on a fine computational grid, MLMC gathers an array of solutions across a hierarchy of grids at coarser scales. This has the effect of offsetting the cost of sampling even if the total number of solutions required may be considerably larger. In our previous work, we demonstrated significant reductions in sampling costs through the two aforementioned strategies: either using sparse grid stochastic collocation in place of direct solution, or employing the multilevel Monte Carlo method with direct solution. In this paper, we combine these two ideas, using a surrogate-enhanced MLMC method to further improve sampling efficiency. Our goal is to show that this hybrid approach further reduces the sampling costs of MLMC while maintaining the accuracy of outcomes.

\subsection{The Grad-Shafranov free boundary problem with uncertainty}

\begin{figure}[!b]\centering
\includegraphics[width=0.46\linewidth]{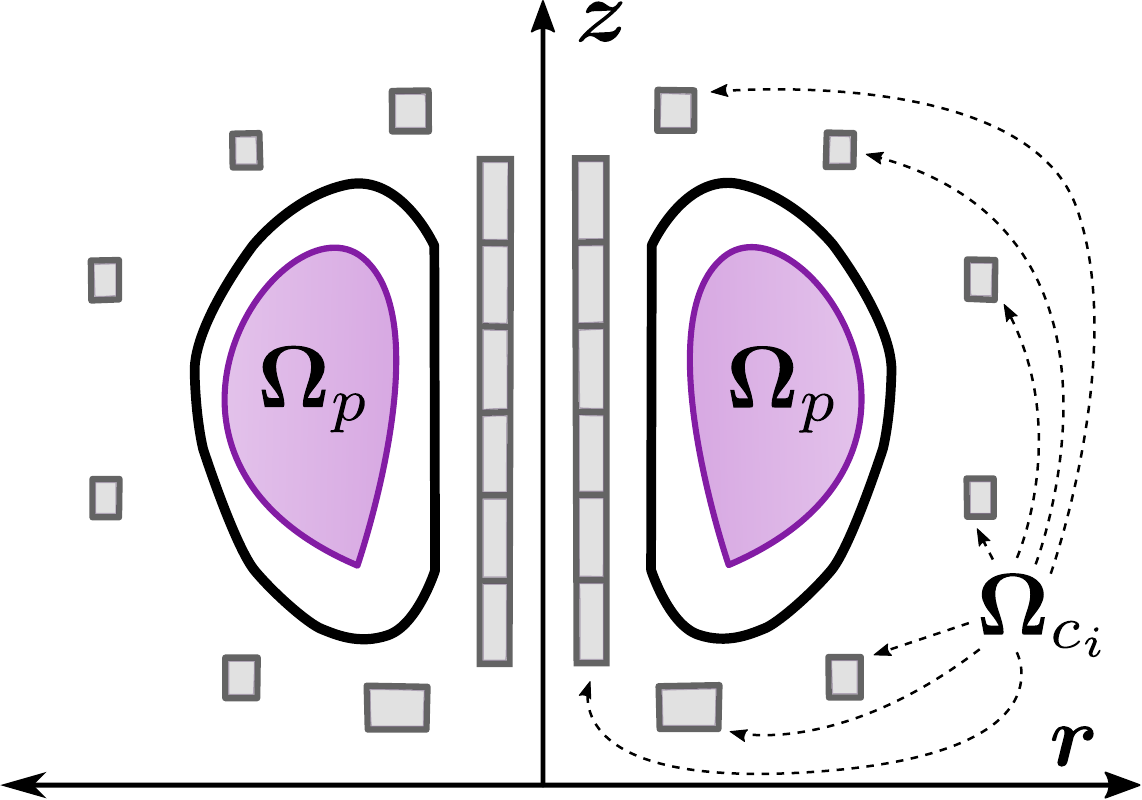}
\caption{Schematic of the cross section of a tokamak. The solid black line represents the wall of the reactor. The grey rectangles represent the coils, each located in a region denoted as $\Omega_{c_i}$, with the index $i$ running over the total number of coils present in the device. The violet region $\Omega_p$ represents the region occupied by the plasma. Due to the problem's (anti)symmetry, it is enough for the analysis to focus on the right half side of the diagram.} 
\label{fig:ReactorScheme} 
\end{figure}
In a magnetic confinement fusion reactor, strong magnetic fields produced by external coils are used to contain a hot plasma in the interior of a vacuum vessel. The confined plasma---a mixture of ionized light atomic nuclei---is heated to induce the merging of the atomic nuclei and the release of a significant amount of energy in a process known as a thermonuclear fusion reaction. During confinement, charged particles in the plasma undergo forces due to the hydrostatic pressure, $p$, and to the magnetic field, $\boldsymbol B$, generated by the currents, $\boldsymbol J$, flowing through the external coils and through the plasma. A state of equilibrium is reached when the magnetic pressure equals the hydrostatic pressure and is expressed mathematically by requiring that the condition ${\boldsymbol J}\times{\boldsymbol B} = \nabla p$ holds in the entire space.

If the reactor possesses axial symmetry, the equilibrium condition can then be stated in terms of a nonlinear partial differential equation for a scalar variable known as the \textit{poloidal flux}, $\psi$. In this study, we will focus on a particular family of axially symmetric reactors known as \textit{tokamaks}. The cross section of the typical configuration is depicted schematically in Figure \ref{fig:ReactorScheme}. A vacuum vessel (solid curved line) contains the plasma (violet region), while an array of coils (solid gray rectangles) generate the confining magnetic field.

Using cylindrical coordinates $(r, z, \varphi)$, this equation, known as the Grad-Shafranov equation \cite{GrRu:1958, LuSc:1957, Shafranov:1958}, is posed in the $(r,z)$ plane and takes the form of
\begin{equation*}
 -\nabla\,\cdot\,\left(\frac{1}{\mu r}\nabla \psi\right) = J_\varphi(\psi,r),
 \end{equation*}
where $\nabla$ and $\nabla\cdot$ denote the Cartesian gradient and divergence operators in two dimensions respectively; the magnetic permeability $\mu$ is either a function of the magnetic field $\mu = \mu(|\nabla \psi|^2/r^2)$ within any ferromagnetic structure, or equal to the constant permeability of vacuum, $\mu_0$, everywhere else. The source $J_\varphi$ accounts for the currents present in the system and is known as \textit{toroidal current density}. This term is defined piecewise and different from zero only in the two following cases:
\begin{itemize}
\item Within a cross section of a coil, it is constant and equal to the ratio between the current going through that coil, $I_i$, and the cross sectional area of the coil $S_i$. \item Within the plasma (confined within the region $\Omega_p$), the current can be represented in terms of the hydrostatic pressure and an additional scalar variable $g$ as
\[
J_\varphi(\psi,r) = r\frac{d}{d\psi} p(\psi) + \frac{1}{2\,\mu r} \frac{d}{d\psi} g^2(\psi).
\]
\end{itemize}
Both the pressure and the scalar function $g$ are considered constant along the streamlines of the magnetic field and thus take the forms $p=p(\psi)$ and $g=g(\psi)$. In this study, following the model proposed in \cite{LuBr:1982}, we adopt the following forms
\begin{equation}\label{eq:source}
\frac{d}{d\psi}p(\psi) = j_0\frac{\beta}{r_0}\left(1-\psi_N^{a_1}\right)^{a_2},  \qquad \text{ and }\qquad
\frac{1}{2}\frac{d}{d\psi}g^2(\psi) = j_0\mu_0r_0(1-\beta)\left(1-\psi_N^{a_1}\right)^{a_2},
\end{equation}
where $\psi_N \in [0,1]$ represents a normalization of $\psi$. The parameter $r_0$ corresponds to the outer radius of the vacuum chamber, while $a_1$ and $a_2$ govern the sharpness of the current peaks near the magnetic axis; $\beta$ referred to as \textit{poloidal beta}, serves as a parameter measuring the ratio between the hydrostatic pressure in the plasma due to temperature (usually called \textit{plasma pressure}) and the pressure attributed to the external magnetic field (or \textit{magnetic pressure}), $\mu_0$ is the magnetic permeability of vacuum, and $j_0$ is introduced as a normalization factor. 
The region $\Omega_p$ circumscribed by the last closed level set of $\psi$ that does not intersect any component of the reactor determines the boundary of the plasma, $\partial\Omega_p$, and thus the confinement region. 
When $\partial\Omega_p$ goes through a saddle point of $\psi$, it is referred to as a 
separatrix, and the saddle point is known as an \textit{x-point}.
Since $\psi$ and $\Omega_p$ are not known a priori this leads to a \textit{free boundary problem}. Moreover, the dependence of the source terms $p(\psi)$ and $g(\psi)$---and possibly even of the magnetic permeability $\mu$---on the solution add further complexity to the Grad-Shafranov free boundary problem. 

In practice, all the parameters of the model are susceptible to uncertainties (arising from imperfect measurements, operational variations,  engineering tolerances, etc.) that introduce stochasticity into the solution $\psi$ and its derived quantities. In this study we will consider that the uncertainties may affect the parameters appearing in the definition of the source term $J_\varphi$; namely, the current intensities of the external coils, and the parameters appearing in \eqref{eq:source}. To model these uncertainties, we introduce a $d$-dimensional random variable $\boldsymbol \omega :=(\omega_1,\ldots,\omega_d)$, whose components are independent and uncorrelated, 
and whose $k$th component $\omega_k$ is associated with the uncertainty in the 
$k$th parameter. We will denote the joint probability density function by $\pi(\boldsymbol\omega)$ and the parameter space by $\boldsymbol W$.

Incorporating the uncertainty, the partial differential equation resulting from the equilibrium condition becomes
\begin{equation}
\label{eq:FreeBoundary}
 -\nabla\,\cdot\,\left(\frac{1}{\mu(\psi,\boldsymbol{\omega}) r}\nabla \psi(\cdot, \boldsymbol{\omega})\right) = \left\{ \begin{array}{ll}
r\frac{d}{d\psi} p(\psi, \boldsymbol{\omega}) + \frac{1}{2\,\mu r} \frac{d}{d\psi} g^2(\psi, \boldsymbol{\omega}) & \quad \text{ in } \Omega_p(\psi, \boldsymbol{\omega})\,, \\
I_k(\boldsymbol \omega)/S_k & \quad \text{ in } \Omega_{W_k} \\
0 & \quad \text{ elsewhere.} 
\end{array}\right.
\end{equation}
Our objective is to devise an efficient computational algorithm that will allow us to quantify the effect that stochasticity in the problem parameters has on the poloidal flux and derived physical properties as measured by an approximation to the expected equilibrium configuration
\begin{equation}
     \label{eq:QoI}
\mathbb{E}\left[\psi(\cdot, \boldsymbol \omega)\right]=\int_{\boldsymbol W} \psi(\cdot,\boldsymbol{\omega})\pi(\boldsymbol\omega)d\boldsymbol{\omega}\,.
\end{equation}
%
\subsection{Overview of the article}\label{sec:Overview}
%
While our motivating example involves the poloidal flux $\psi$, in order to make the subsequent analysis applicable to a broader class of problems, we introduce a generic state variable $u$ to represent the quantity of interest. In its simplest form, a Monte Carlo estimate of the expectation $\mathbb E[u]$ gathers a large number, $N$, of the realizations of $u(\cdot, \boldsymbol\omega)$ and estimates the expectation by the mean of the sample as
\[
\mathbb E[u] \approx \frac{1}{N}\sum_{i=1}^N u\left(\cdot, \boldsymbol\omega^{(i)}\right)\,.
\]
Based on the sample mean, the approximation is unbiased and, by the central limit theorem, it will converge to the true value as $N^{-1/2}$. Due to the slow rate of convergence, a large number of samples may need to be gathered to obtain a reasonably accurate approximation of $\mathbb E[u]$. Therefore, in simple words, the cost of obtaining a Monte Carlo estimate boils down to the workload associated with sampling.
In our case, $u$ maps the stochastic parameters in \eqref{eq:FreeBoundary} to the solution of the free boundary problem. In practical terms, the task of obtaining a Monte Carlo approximation for $\mathbb E[u]$ entails the use of a computational discretization of the mapping, and the computation of a large number of numerical solutions to \eqref{eq:FreeBoundary} for different values of the stochastic parameters. Unfortunately, even if obtaining such a pool of numerical solutions is computationally feasible, the cost of numerically solving the free boundary problem repeatedly can make the process unpractical. The goal then is then to bypass the need for a direct numerical solution of \eqref{eq:FreeBoundary} for every new sample gathered, or at least mitigate the work required to obtain each sample.

In this study, we explore ways to reduce costs using \textit{surrogates}. A surrogate of $u$ is a function $\hat u$ that---at the expense of a certain degree of accuracy---circumvents the computational expense incurred by a direct evaluation of $u$. There are two aspects of using surrogates, the so-called \textit{online step}, in which the surrogate is constructed, and the \textit{offline computations}, in which the surrogate is used in place of direct evaluation within a Monte Carlo simulation. There is overhead associated with the online step, but if many surrogate evaluations are required for the simulation and the costs of using the surrogate are sufficiently lower than those of direct evaluation, the one-time cost of construction will not be significant. An overview of the paper is as follows. In Section \ref{sec:Surrogate_Construction}, we explore different strategies for building surrogates using stochastic collocation in parameter space, and we analyze the costs of these constructions in an idealized setting where the solution is assumed to have a certain amount of regularity. In Section \ref{sec:Sampling_surrog}, we analyze the costs of sampling using these surrogates in the same idealized setting. The cost analyses of these sections build on methods used to analyze costs in \cite{ClGiScTe:2011,TeScGiUl:2013}, adapted to account for the use of surrogate approximations. Finally, in Section \ref{sec:Num-Exp}, we report on the experimental performance of the surrogates for use with the Grad-Shafranov problem. This problem is not as smooth as needed for the cost analysis, but the experiments show a significant reduction in costs and suggest that even in this more complex setting, the cost analysis gives insight into performance.

\section{Construction of the surrogate function}
\label{sec:Surrogate_Construction}
%
\subsection{Sparse grid stochastic collocation}\label{sec:SC}

We now present a brief overview of the sparse grid stochastic collocation method \cite{BaNoRi:2000, KlBa:2005, MaNi:2009, Sm:1963} for approximating the solution to \eqref{eq:FreeBoundary} with stochastic parameters that will be denoted by $x$ and, for simplicity, will be assumed to be restricted to the $d$-dimensional unit cube $[0,1]^d$. Assume that for every fixed realization of the parameter $\boldsymbol \omega\in \boldsymbol W$, $u(\cdot,\boldsymbol \omega)$ is an element of a function space $Z$. We wish to build an interpolatory approximation of $u(\cdot, \boldsymbol \omega)$ based on a set of observations at particular values of $\boldsymbol \omega$. The stochastic collocation method uses particular values of $\boldsymbol \omega$ that give rise to a nested family of grids in the parameter space, where each grid consists of $m_i$ values of $\boldsymbol \omega$ that we shall call \textit{nodes}.

To mitigate the so-called \textit{curse of dimensionality}, the sequence of grids can be built by choosing the nodes according to Smolyak's rule, which curtails the growth in the number of nodes involved---although at the expense of some accuracy---and leads to a sequence of grids known as \textit{sparse grids}. The construction process can be summarized as follows. Let
\[
W^1:=\left\{\omega_1^1,\ldots, \omega_{m_1}^1\right\}\subset \cdots\subset W^{i}: = \left\{\omega_1^i,\ldots, \omega_{m_i}^i\right\} \subset W^{i+1}: = \left\{ \omega_1^{i+1},\ldots,  \omega_{m_{i+1}}^{i+1}\right\}\subset \cdots
\]
be a family of points in the unit interval $[0,1]$ (typically associated with a univariate quadrature rule; here, the superscript of a node denotes the grid (or level) to which it belongs and the subscript corresponds to an ordering within each level. These can be used to define a sparse grid $H(q,d)$ of dimension $d$ and level $q$ by 
\begin{equation}
\label{eq:NestedColPts}
H(q,d) = \bigcup_{q-d+1\le|\boldsymbol{i}|\le q} \left(W^{i_1}\times \cdots\times W^{i_d}\right)\in [0,1]^d, 
\end{equation}
where $|\boldsymbol{i}| = i_1+\cdots+i_d$. This set of points can then be used to interpolate the function $u$ using Smolyak's interpolant of level $q$, defined by 
\begin{equation}
\label{eq: Smolyak_Quad_formula}
\widehat u:= \sum_{q+1\le |\boldsymbol{i}|\le q+d} (-1)^{q+d-|\boldsymbol{i}|} \binom{d-1}{q+d-|\boldsymbol{i}|}\cdot \left(\mathrm I_{W^{i_1}}\otimes\cdots\otimes \mathrm I_{W^{i_d}}\right) (u).
\end{equation} 
where
\[
\mathrm I_{W^{i}}(u)(\cdot,\boldsymbol{\omega}):=\sum_{j=1}^{m_{i}} u\left(\cdot,\omega_j^i\right)\,\phi_j(\boldsymbol{\omega})
\]
is a univariate interpolation operator on the set of points $W^{i}$, and for $\omega_j^i\in W^i$ the interpolating basis functions $\phi_k$ satisfy the Kronecker property; i.e. $\phi_k(\omega_j^i)= 1$ if $k=j$ and $\phi_k(\omega_j^i)=0$  if $k\neq j$.

The approximation properties of the interpolant will depend strongly on the regularity of the mapping $u$ as a function of the parameters. For instance, if for some $\boldsymbol\omega = (\omega_1\ldots,\omega_d)\in \boldsymbol W$ we have $u \in C^0(\boldsymbol W,Z)$ and for any $\omega_k$ the function $u(\cdot,\omega_k)$ obtained from $u(\cdot,\boldsymbol\omega)$ by freezing all coordinates of $\boldsymbol\omega$ with the exception of $\omega_k$ admits an analytic extension in a neighborhood of $\boldsymbol\omega$, then it is possible \cite{NoTeWe:2008,TeJaWe:2015} to show that
\begin{equation}
\label{eq:coll-error-bound-1}
  \big\|u-\widehat u\,\big\|_\infty \leq C P^{-\nu},
\end{equation}
where $P$ denotes the number of nodes in the sparse grid, $C$ is a positive constant and the power $\nu$ is an increasing function of the size of the domain of definition of the function's analytic extension in the complex plane. On the other hand, if the mapping is less regular with respect to the stochastic parameter $\boldsymbol\omega$, for instance of class $C^k(\boldsymbol W,Z)$, then \cite[Theorem~8]{BaNoRi:2000}
\begin{equation}
\label{eq:coll-error-bound-2}
  \big\|u-\widehat u\,\big\|_\infty \leq C P^{-k} |\log P|^{(k+2)(d-1)+1}\,. 
\end{equation}
In the asymptotic regime as the number of sparse grid nodes $P$ grows, this estimate is dominated by the factor $P^{-k}$. In analyzing costs, we will assume that $P$ is sufficiently large for an estimate of the form \eqref{eq:coll-error-bound-1} to hold.

\subsection{Notational considerations}

Following the analysis from \cite{Gr:1999}, if $\Omega$ is a bounded Lipschitz domain fully enclosing the component structures of the reactor and all the external coils $\Omega_{c_i}$, we will define the solution space to the free boundary problem \eqref{eq:FreeBoundary} as
\[
    Z:=\overline{\left\{f:\Omega\rightarrow \mathbb{R} \,\Bigg| \,\int_\Omega f^2rdrdz<\infty; \,  \int_\Omega\frac{|\nabla f|^2}{r}drdz<\infty; \, f(0,z)=0 \right\}\cap C^0(\overline{\Omega})}\,.
\]
Above, the closure is taken with respect to the inner product and the energy norm
\[
    \langle f,g\rangle := \int_{\Omega} \frac{1}{r}\; \nabla f\cdot\nabla g \;drdz,\qquad \qquad \| f \|_{Z} :=\left(\int_\Omega\frac{|\nabla f|^2}{r} \;drdz\right)^{1/2}.
\]
The mapping $u: \boldsymbol W \to Z$ denotes the solution operator mapping a realization of the random variable $\boldsymbol \omega$ to the corresponding solution of \eqref{eq:FreeBoundary}. Our goal is to use the sparse grid stochastic collocation method to construct a surrogate for this solution operator. We will do so by computing the discrete solution of \eqref{eq:FreeBoundary} for the values of $\boldsymbol\omega$ corresponding to  Clenshaw-Curtis quadrature nodes \cite{BaNoRi:2000,ClCu:1960} in the sparse grid \eqref{eq:NestedColPts} and using these solutions to build an interpolant as given by \eqref{eq: Smolyak_Quad_formula}. 

To measure distances between two random variables $u,v: \boldsymbol W\to Z$ we will use the metric 
\[
\left\Vert u - v \right\Vert_{L^2(\boldsymbol W,Z)} =
    \left(\int_{\boldsymbol W} \left\Vert u(\cdot,\boldsymbol{\omega}) - v(\cdot,\boldsymbol{\omega}) \right\Vert_{Z}^2 d\mathbb{P}(\boldsymbol{\omega}) \right)^{1/2} = \left(\mathbb{E}\left[\left\Vert u(\cdot,\boldsymbol{\omega}) - v(\cdot,\boldsymbol{\omega}) \right\Vert_{Z}^2\right]\right)^{1/2}\,,
\]
where the probability density function $d\mathbb P(\boldsymbol\omega)$ corresponds to the measure in the parameter space $\boldsymbol W$. The quantity
\begin{equation}
\label{eq:Variance}
\mathbb{V}[u] := \mathbb{E}\left[\left\Vert u - \mathbb{E}[u]\right\Vert_Z^2\right]\,,
\end{equation}
which shares many properties and similarities with the variance, will be a useful measure of the spread of the surrogate evaluations and thus will become a useful tool for quantifying statistical errors.

In what follows, $u_h$ will represent a discrete approximation of $u$ obtained using a grid in physical space characterized by the mesh parameter $h$ and consisting of $M$ nodes; the Smolyak surrogate for $u_h$ of level $q$ defined by \eqref{eq: Smolyak_Quad_formula} will be denoted by $\widehat u_{h}$. We will also assume that the sample-wise discretization error and interpolation error are bounded as

\begin{subequations}
\label{eq:Assumption_uh}
\noindent\begin{minipage}{.5\linewidth}
\begin{equation} \label{eq:Assumption_uhA}
\left\|u\left(\cdot,\boldsymbol\omega^{(i)}\right)-u_h\left(\cdot,\boldsymbol\omega^{(i)}\right)\right\|_Z\leq C_m\left(\boldsymbol\omega^{(i)}\right)M^{-\alpha}\,,
\end{equation}
\end{minipage}%
\begin{minipage}{.5\linewidth}
\begin{equation}\label{eq:Assumption_uhB}
\left\|u_h\left(\cdot, \boldsymbol\omega^{(i)}\right)-\widehat u_{h}\left(\cdot, \boldsymbol\omega^{(i)}\right)\right\|_Z\leq C_p\left(\boldsymbol\omega^{(i)}\right)P^{-\nu}\,,
\end{equation}
\end{minipage}
\end{subequations}
\vspace{.1cm}

\noindent where $\alpha$ and $\nu$ are the orders of the sample-wise discretization error and interpolation error in \eqref{eq:coll-error-bound-1}. The exponent $\alpha$ depends on both the spatial dimension $d$ and the choice of norm for $Z$. In the two--dimensional case that we are dealing with, the number of spatial grid nodes, $M$, behaves like $M \propto h^{-2}$. Therefore, under the usual (and reasonable) assumption of $H^2$--regularity for $u$, the discretization achieves $\mathcal{O}(h^2)$ convergence in the $L^2$ norm and $\alpha = 1$ \cite[Corollary 7.7]{Braess2007}.

The constants $C_m(\boldsymbol\omega^{(i)})$ and $C_p(\boldsymbol\omega^{(i)})$ depend only on the problem geometry and the particular realization $\boldsymbol\omega^{(i)}$. Under the assumption that the solution's regularity is uniformly controlled across samples, they are both bounded uniformly by deterministic constants $C_m$ and $C_p$. Specifically, the assumption that the solution's derivatives satisfy uniform bounds almost surely \cite{BaNoTe:2007,BaScZo:2011,NoTeWe:2008,TeJaWe:2015,TeScGiUl:2013} ensures the finiteness of the bounds 
\[
C_m := \sup_{\boldsymbol{\omega} \in \Omega} C_m(\boldsymbol{\omega}) \qquad \text{ and } \qquad C_p := \sup_{\boldsymbol{\omega} \in \Omega} C_p(\boldsymbol{\omega})\,.
\]
We remark that, in what follows, when describing the different approaches to surrogate building, the associated values of the constants $C_m$, $C_p$, $\alpha$, and $\nu$ may vary from section to section but their properties, as enlisted above, remain true.
\subsection{Collocation with a single level of spatial discretization }
\label{sec: SLSGC-SL}
We start by exploring the simplest case: a sparse-grid based surrogate constructed based on samples of $u_h$ obtained using a single discretization mesh in physical space. The total error incurred by the surrogate approximation $\widehat u_h$ can be split into two components
\begin{equation}
\label{eq:SLSGC_SL_Err_exact_surrog}
    \left\Vert u - \widehat u_{h}\right\Vert_{L^2(\boldsymbol W,Z)}\le \|u-u_h\|_{L^2(\boldsymbol W,Z)}+ \|u_h-\widehat u_{h}\|_{L^2(\boldsymbol W,Z)},
\end{equation}
where the first term accounts for the discretization error, and the second one reflects the interpolation error. Using the estimate above together with the sample-wise error estimates \eqref{eq:Assumption_uh}, we will now determine the number of spatial and sparse grid nodes required to achieve an approximation error below a certain user-specified tolerance $\epsilon$.

Let us introduce two numbers $\theta_1, \theta_2 \in (0,1)$ such that $\theta_1+\theta_2 = 1$. 
We will use these numbers, known as the {\it splitting ratios}, to dictate the maximum 
allowable contribution of each type of error to the total error budget by requiring that 
\begin{equation} \label{eq:error-budget}
\|u-u_h\|_{L^2(\boldsymbol W,Z)}\le C_mM^{-\alpha}\le \theta_1\epsilon \qquad\text{ and }\qquad \|u_h-\widehat u_{h}\|_{L^2(\boldsymbol W,Z)} \le C_{p} P^{-\nu}\le \theta_2\epsilon\,.  
\end{equation}
We will defer the discussion of the strategy for selecting suitable values for $\theta_1$ and $\theta_2$ to Section \ref{sec:Sampling_surrog}. Using the conditions above, the number of physical and sparse grid nodes $M$ and $P$ required to obtain the desired tolerance can be bounded as
\begin{equation}
\label{eq:SLSGC_SL_SpatialGridsNo_n_SparseGridsNo}
M\ge \left(\frac{\theta_1\epsilon}{C_m}\right)^{-\frac 1 {\alpha}} \quad \text{ and } 
\quad P\ge \left(\frac{\theta_2\epsilon}{C_p}\right)^{-\frac{1}{\nu}}. 
\end{equation}
The sparse and discretization grids used in a practical simulation should be the ones with the smallest number 
of nodes that satisfy \eqref{eq:SLSGC_SL_SpatialGridsNo_n_SparseGridsNo}.
These estimates also suggest that the number of sparse grid nodes $P$ can be made a function of the number of spatial grid nodes $M$ through the common parameter $\epsilon$. Considering equality in both expressions above and eliminating the tolerance from the system leads to
\[
\frac{P^\nu}{C_p} = \frac{\theta_1}{\theta_2} \frac{M^\alpha}{C_m}
\]
Substituting this expression for $P^\nu$ into the interpolation error estimate and 
using \eqref{eq:SLSGC_SL_Err_exact_surrog} yields
\begin{equation} \label{eq:error-surrogate}
 \|u-{\widehat u}_h\|_{L^2(\boldsymbol W,Z)}\le 
C_m \left(1+\frac{\theta_2}{\theta_1}\right) M^{-\alpha},
\end{equation}
which indicates that the error of the surrogate is of essentially the same form as that of the
discrete solution.

Considering that, for a nonlinear problem like the one in our application, the computational cost of obtaining a sample is itself a random variable, we will estimate the computational work required to build this kind of surrogate, by assuming that the \textit{average} work per sample is of the form $W\simeq M^\gamma$ for some $\gamma>0$ that depends on the particular method used to solve the system of equations arising from the discretization,  and $A\simeq B$ means $C_1 B\le A\le C_2 B$ for positive $A$ and $B$, with constants $C_1, C_2$ independent of sample size $N$ and number of spatial grid nodes $M$.
Hence, making use of \eqref{eq:SLSGC_SL_SpatialGridsNo_n_SparseGridsNo},
for a surrogate built using $P$ sparse grid nodes, the expected work to construct the surrogate, measured in terms of $\epsilon$, is given by 
\[
\mathcal{W}_\text{SL-SL}^\text{off}  = PW 
\simeq \epsilon^{-\frac{1}{\nu}}\cdot \epsilon^{-\frac{\gamma}{\alpha}} = \epsilon ^{-\frac 1 {\nu}-\frac{\gamma}{\alpha}},
\]
where the subscript  ``SL-SL'' denotes the surrogate built with a single-level sparse grid and a single-level spatial grid, the superscript ``off'' represents an offline process, and, to recall, $\epsilon$ represents the desired accuracy,  
$\alpha$ is the convergence rate of the spatial discretization,
$\nu$ is the convergence rate of the interpolant, 
$\gamma$ is the rate associated with the 
non-linear solver that characterizes the average work per sample.
%
\subsection{Collocation with multilevel spatial discretization}
\label{sec: SLSGC-ML}
%
Multilevel Monte Carlo sampling \cite{BaScZo:2011,ClGiScTe:2011, Gi:2008, Gi:2015,NoTe:2015,TeScGiUl:2013} is a well-established technique for reducing the online computational cost when sampling involves the numerical solution of a differential equation. The idea is simple to describe: gather as many samples as possible in coarser discretization grids---where the computational cost of solving a differential equation is smaller---and then ``correct" by including a smaller number of samples obtained using finer grids. In this section we take that idea a step further, by replacing the direct solver on each spatial discretization mesh with a sparse grid surrogate built with a fixed accuracy level in parameter space.

Let $\{\mathcal{T}_\ell\}$ be a family of spatial meshes of increasingly higher resolution characterized by the \textit{level} indexed by $\ell =0, \ldots, L$. These meshes will have an increasing number of nodes $\{M_\ell,\}_{0\le \ell \le L}$ that, we will assume, satisfy
\begin{equation}
\label{eq:MeshGrowth}
M_\ell = s M_{\ell-1} \qquad \text{ for } s>1.
\end{equation}
This assumption will hold asymptotically, for instance, for uniformly refined meshes. Let $u_\ell$ be the discrete approximation of $u$ on $\mathcal{T}_\ell$, and $\widehat u_\ell$ be the surrogate obtained by gathering samples of $u_\ell$ for parameter values corresponding to the $P$ nodes of a sparse grid of level $q$ as prescribed by \eqref{eq: Smolyak_Quad_formula}. The multilevel approach will then require the construction of one surrogate for each level of spatial discretization, resulting in a collection $\left\{\,\widehat u_\ell\right\}_{\ell=0}^L$. The surrogate corresponding to the finest spatial discretization level $L$ can be expressed as a telescoping sum involving surrogates built at coarser discretization levels as
\begin{equation}
    \label{eq:SLSGC-ML_surrogate}
    \widehat{u}_L = \sum_{\ell=0}^L \widehat u_{\ell} -\widehat u_{\ell-1} = \sum_{\ell=0}^L \widehat Y_{\ell}, \qquad \text{ where } \qquad Y_\ell := 
    \begin{cases}
    u_\ell-u_{\ell-1} & \text{ for } \ell\ge 1\,,\\
    u_0 & \text{ for } \ell = 0\,.
    \end{cases}
\end{equation}
The terms $Y_\ell$ can be considered as a sequence of successive corrections to the initial discretization $u_0$ on two consecutive spatial mesh levels, leading to the finest discretization $u_\ell$. Thus, in view of the linearity of the interpolation operator $\widehat Y_{\ell} := \widehat u_{\ell}-\widehat u_{\ell-1}$ for $\ell\ge 1$ and $\widehat Y_{0}:=\widehat u_{0}$ can be considered as corrections to the surrogate $\widehat u_0$ built using the same sparse grid on adjacent spatial meshes $\mathcal{T}_{\ell-1}$ and $\mathcal{T}_{\ell}$. 

For a fixed realization $\boldsymbol \omega^{(i)}$ of the random variable, an intuitive reading of the difference 
\[
\left\|Y_\ell^{(i)} - \widehat Y_{\ell}^{(i)}\right\|_Z = \left\|\left(u_\ell^{(i)} - u_{\ell-1}^{(i)}\right) - 
\left(\widehat{u}_\ell^{(i)} - \widehat{u}_{\ell-1}^{(i)}\right) \right\|_Z = \left\|\left(u_\ell^{(i)} - \widehat u_\ell^{(i)}\right) - \left( u_{\ell-1}^{(i)} - \widehat u_{\ell-1}^{(i)}\right)\right\|_Z 
\]
suggests that:
\begin{enumerate}
\item As spatial discretization improves, it must follow that $\left(u_\ell^{(i)} - u_{\ell-1}^{(i)}\right) \to 0$ and thus  $\left(\widehat{u}_\ell^{(i)} - \widehat{u}_{\ell-1}^{(i)}\right) \to 0$. This would imply that $\left\|Y_\ell^{(i)} - \widehat Y_{\ell}^{(i)}\right\|_Z\to 0$ as $\ell\to\infty$ \textit{even for a fixed number of interpolation points} $P$.
\item As the quality of the interpolation grid improves, both $\left(u_\ell^{(i)} - \widehat u_\ell^{(i)}\right)\to 0$ and $\left(u_{\ell-1}^{(i)} - \widehat u_{\ell-1}^{(i)}\right)\to 0$ implying that $\left\|Y_\ell^{(i)} - \widehat Y_{\ell}^{(i)}\right\|_Z\to 0$ as the number of interpolation nodes $P\to\infty$ \textit{even for a fixed number of discretization points} $M_\ell$. 
\end{enumerate}
These intuitive observations in fact hold for the case of Lagrangian interpolation and Smolyak grids based on nested quadrature points---which include the Clenshaw-Curtis strategy we employ in this paper. Moreover, it was shown in \cite[Section 5]{TeJaWe:2015} that there are in fact positive constants $C^{(i)}, \rho$, and $\nu$ such that
\begin{equation}
\label{eq:Assumption_Yl}
    \left\Vert Y_\ell ^{(i)}-\widehat Y^{(i)}_{\ell}\right\Vert_Z\leq C_3^{(i)}M_\ell^{-\rho}P^{-\nu} \quad \text{for }\;\; \ell\ge 0.
\end{equation}
With the aid of this estimate, an estimate for the interpolation error can be obtained as follows:
\begin{align}
\label{eq:SLSGC_ML_Err_exact_surrog}
\nonumber
    \left\Vert u - \widehat u_{L}\right\Vert_{L^2(\boldsymbol W,Z)} \le\,& \|u-u_L\|_{L^2(\boldsymbol W,Z)}+ \|u_L-\widehat u_L\|_{L^2(\boldsymbol W,Z)} \\
    \nonumber
    \le\,&  \|u-u_L\|_{L^2(\boldsymbol W,Z)}+ \sum_{\ell=0}^L \left\Vert  Y_{\ell}-\widehat Y_{\ell}\right\Vert_{L^2(\boldsymbol W,Z)} &\qquad {\text{\small(Telescopic expansion for $u_L$ and $\widehat u_L$ )}},\\
    \le\,& C_m M_L^{-\alpha}+C_pP^{-\nu} \sum_{\ell=0}^L M_\ell^{-\rho} &\qquad {\text{\small(Using \eqref{eq:Assumption_uhA} and \eqref{eq:Assumption_Yl})}}.
\end{align}
To ensure that the discretization error falls below $\theta_1\epsilon$, we estimate the number of points on the finest grid $M_L$ and the required spatial grid level $L$ as 
\begin{equation}
    \label{eq:SLSGC_MLS_SpatialGridsNo}
    M_L = M_0s^L \ge \left(\frac{\theta_1\epsilon}{C_m}\right)^{-\frac 1 {\alpha}} \qquad \text{ and } \qquad     L = \left\lceil \frac{1}{\alpha}\log_s \left(\frac{C_m}{\theta_1M_0^\alpha\epsilon}\right) \right\rceil,
\end{equation}
where $\left\lceil\cdot\right\rceil$ denotes the ceiling function.

We now estimate an upper bound for the interpolation error in terms of the number of interpolation nodes, $P$. 
Using  \eqref{eq:Assumption_Yl} 
and enforcing that the interpolation error remains below the tolerance $\theta_2\epsilon$, we obtain
\begin{align*}
   \|u_L-\widehat u_{L}\|_{L^2(\boldsymbol W,Z)}&\le C_pP^{-\nu} \sum_{\ell=0}^L M_\ell^{-\rho}= C_pP^{-\nu}M_0^{-\rho}\sum_{\ell=0}^L s^{-\rho \ell}\le P^{-\nu}\frac{C_pM_0^{-\rho}}{1-s^{-\rho}}\le\theta_2\epsilon.
\end{align*}
This inequality leads to an estimate for the number of required grid points $P$ in the sparse grid
\begin{equation}
   \label{eq:SLSGC_ML_SparseGridsNo}
    P\ge\left(\frac{\left(1-s^{-\rho}\right)\theta_2\epsilon }{C_pM_0^{-\rho}}\right)^{-\frac{1}{\nu}} = \left(\frac{\theta_2 C_m (1-s^{-\rho})}{\theta_1 C_pM_0^{-\rho}}\right)^{-1/\nu}M_L^{\alpha/\nu} 
     \simeq M_L^{\alpha/\nu}.
\end{equation}
Recalling that $\nu$ is the rate of convergence of the interpolant and $\alpha$ is that of spatial discretization, the estimate above tells us that the growth in the number of required interpolation points as a function of the number of discretization points is driven by the relative space vs. parameter regularity of $u$. When the ratio is balanced, the growth in the nodes will be approximately linear; if the mapping is smoother in parameter space than it is in physical space then $\alpha/\nu<1$ and few interpolation nodes will be enough, while if the mapping is smoother in space then $\alpha/\nu>1$ and more interpolation points will be necessary to catch up with the discretization accuracy. 

The work associated with the construction of the surrogate depends on several factors: the number of nodes on the sparse grid in parameter space, the number of spatial discretization levels used, the number of nodes on each spatial mesh, and the average work required for a direct solve on each of the spatial grids (i.e. the work associated with a direct computation on each discretization level). Assuming that the average work per solve on the $\ell$-th spatial discretization level is 
$W_\ell \simeq M_{\ell}^{\gamma}$, using $M_{\ell}=s^{\ell}M_0$, and 
recalling that equation \eqref{eq:SLSGC_MLS_SpatialGridsNo} implies that $L \propto -\frac{1}{\alpha}\log_s\epsilon$, we can estimate the total work by adding across discretization levels as
\[
    \sum_{\ell=0}^L W_\ell \simeq \sum_{\ell=0}^L {M_\ell^\gamma} 
     =  M_0^\gamma \left(\sum_{\ell=0}^L s^{\gamma \ell}\right) 
     = M_0^\gamma \left(\frac{s^{\gamma(L+1)}-1}{s^\gamma-1}\right)
     \simeq  M_0^\gamma  s^{\gamma L} 
     \simeq  s^{-\frac{\gamma}{\alpha} \log_s\epsilon} 
     =  \epsilon^{-\gamma/\alpha}.
\]
This estimate, together with \eqref{eq:SLSGC_SL_SpatialGridsNo_n_SparseGridsNo}, yields the total work required to build this surrogate as
\begin{equation}
    \label{eq:SLSGC_MLS_Cost_Construct}
    \mathcal{W}_\text{SL-ML}^\text{off} = P\sum_{\ell=0}^L W_\ell \,\simeq \epsilon^{-1/\nu} \cdot \epsilon^{-\gamma/\alpha} = \epsilon^{-\frac 1 {\nu}-\frac{\gamma}{\alpha}},
\end{equation}
where the subscript ``SL-ML'' denotes the surrogate built with a single-level sparse grid and a multilevel spatial discretization, and again the superscript ``off'' denotes the work done offline.
Since the most expensive computations required in surrogate construction take place at the finest grid level $L$, the asymptotic estimate above coincides with the one for the single-level surrogate discussed in Section \ref{sec: SLSGC-SL}. We remark, however, that given the fact that for this strategy one must construct a surrogate for each discretization level $\ell$ the offline work will be strictly larger than for the single-level strategy. 

\section{Monte Carlo estimation and sampling costs}
\label{sec:Sampling_surrog}

We now turn to the \textit{online} costs of the surrogate-based estimation. 
Concretely, we will address the issues of efficiency and accuracy of the surrogates obtained from the sparse grid collocation developed in Section \ref{sec:Surrogate_Construction}.

\subsection{Sampling with a single-spatial level surrogate}

We start with the simplest Monte Carlo estimation for the expectation described in Section \ref{sec:Overview}, where evaluations of the mapping $u(\boldsymbol\omega)$ are replaced by evaluations of the single-level discrete surrogate $\widehat u_h(\boldsymbol\omega)$ described in Section \ref{sec: SLSGC-SL}. When no surrogate is used and the discretization of $u_h$ is obtained through the finite element method, this technique is often referred to as the Finite Element Monte Carlo method. 

Let $\widehat u_{h}^{(i)}$ denote the evaluation of the surrogate for the $i$-th realization of the parameter $\boldsymbol{\omega}$. The surrogate-based Monte Carlo estimator for the mean of $u$ is defined as the sample mean
\[
    A\left[\widehat u_{h}\right] := \frac{1}{N}\sum_{i=1}^{N} \widehat u_{h}^{(i)}.
\]
This estimator is unbiased and verifies the properties
\begin{equation}
\label{eq:MCproperties}
\mathbb{E}[A[\widehat u_h]] = \mathbb{E}[\widehat u_{h}] \qquad \text{ and } \qquad \mathbb{V}[A[\widehat u_h]] = \mathbb{V}[\widehat u_{h}]/N\,,
\end{equation}
where $\mathbb V[\cdot]$ is defined as in \eqref{eq:Variance}. To assess the accuracy of the estimator we will use the \textit{mean squared error} (MSE) defined as
 \[
\mathcal{E}_{A}^2:=\mathbb E\left[\left\Vert\mathbb{E}[u]-A\left[\widehat u_{h}\right] \right\Vert_{Z}^2\right]\,
\] 
which, using \eqref{eq:MCproperties}, can be decomposed into two terms: one due to bias, $\mathcal{E}_{\text{Bias}}$, and one due to variance, $\mathcal{E}_{\text{Stat}}$, as follows:
\[
\mathcal{E}_{A}^2 =\mathbb E \left[ \left\Vert\mathbb{E}[u]-\mathbb{E}\left[\widehat u_{h}\right]\right\Vert^2_Z 
 \right]+ \mathbb E \left[ \left\Vert \mathbb{E}\left[\widehat u_{h}\right] - A\left[\widehat u_{h}\right]\right\Vert_Z^2\right]=\left\Vert\mathbb{E}[u]-\mathbb{E}\left[\widehat u_{h}\right]\right\Vert^2_Z + \frac{\mathbb{V}\left[\widehat u_{h}\right]}{N}=\mathcal{E}_{\text{Bias}}^2 + \mathcal{E}_{\text{Stat}}^2,
 \]
 where we have implicitly defined
 \[
\mathcal{E}_{\text{Bias}}^2 : = \left\Vert\mathbb{E}[u]-\mathbb{E}\left[\widehat u_{h}\right]\right\Vert^2_Z \qquad \text{ and } \qquad \mathcal{E}_{\text{Stat}}^2:= \frac{\mathbb{V}\left[\widehat u_{h}\right]}{N}.
\]
The bias term can be further bounded by the sum of discretization error, $\mathcal{E}_{\text{Dis}}:=\|\mathbb E [u] - \mathbb E [u_h]\|_Z$, and interpolation error \,$\mathcal{E}_{\text{Interp}}:= \left\Vert\mathbb E [u_h]-\mathbb E \left[\widehat u_{h}\right]\right\Vert_Z$, as
\[
\mathcal{E}_{\text{Bias}} = \left\Vert\mathbb{E}[u]-\mathbb{E}\left[\widehat u_{h}\right]\right\Vert_Z \leq \|\mathbb E [u] - \mathbb E [u_h]\|_Z + \left\Vert\mathbb E [u_h]-\mathbb E \left[\widehat u_{h}\right]\right\Vert_Z = \mathcal{E} _{\text{Dis}} + \mathcal{E}_{\text{Interp}}\,.
\]
Putting all these together leads to the estimate
\[
\mathcal{E}_{A}^2 \leq \left(\mathcal{E} _{\text{Dis}}+ \mathcal{E}_{\text{Interp}}\right)^2 + \mathcal{E}_{\text{Stat}}^2.
\]
If the expectation of $u$ is nonzero, the \textit{normalized mean squared error}  (nMSE), denoted $\overline{\mathcal{E}}_{A}^2$, is considered. A similar normalization and notation is used for the discretization, interpolation and statistical errors. When the normalizing factor $\left\Vert\mathbb{E}[u] \right\Vert_{Z}^2$ is not available or computable we approximate it by $\left\Vert\mathbb{E}[u_h] \right\Vert_{Z}^2$. For instance the nMSE is approximated as
\[
\overline{\mathcal{E}}_{A}^2 \approx \frac{\left\Vert\mathbb{E}[u]-\mathbb{E}\left[\widehat u_{h}\right]\right\Vert^2_Z}{\left\Vert\mathbb{E}[u_h] \right\Vert_{Z}^2} + \frac{\mathbb{V}\left[\widehat u_{h}\right]}{N\left\Vert\mathbb{E}[u_h] \right\Vert_{Z}^2}\le \left(\overline{\mathcal{E}} _{\text{Dis}}+ \overline{\mathcal{E}}_{\text{Interp}}\right)^2 + \overline{\mathcal{E}}_{\text{Stat}}^2.
\]
Given a target tolerance $\epsilon^2$ for the nMSE, the contribution of these three errors can be controlled by requiring that  
\begin{equation}
\label{eq:MC_nMSE_ErrSplit}
\overline{\mathcal{E}}_{\text{Dis}} \le \theta_1\epsilon, \qquad 
\overline{\mathcal{E}}_{\text{Interp}} \le \theta_2 \epsilon,\qquad  \text{ and } \qquad
\overline{\mathcal{E}}_{\text{Stat}} = \sqrt{\frac{\mathbb{V}\left[\widehat u_{h}\right]}{N\left\Vert\mathbb{E}[u_h] \right\Vert_{Z}^2}} \le \sqrt{\theta} \epsilon,    
\end{equation}
where $\theta\in(0,1)$ is the splitting parameter between the relative bias and the statistical errors, and $\theta_1$ and $\theta_2$ are the splitting parameters between the discretization and interpolation errors, such that $\theta_2= \sqrt{1-\theta} - \theta_1$. The number of grid points in the physical space $M$ and the parameter space $P$ are determined by \eqref{eq:SLSGC_SL_SpatialGridsNo_n_SparseGridsNo} with a scaling factor $\left\Vert\mathbb{E}[u_h] \right\Vert_{Z}$, while the sample size $N$ is estimated from \eqref{eq:MC_nMSE_ErrSplit} as
\begin{equation}
\label{eq:MC_SampleSize}
         N =  \left\lceil \frac{\mathbb{V}\left[\widehat u_{h}\right]}{\theta \epsilon^2\left\Vert\mathbb{E}[u_h] \right\Vert_{Z}^2}\right\rceil 
         \simeq \epsilon^{-2}.
\end{equation}
The average work, $W^e$, required to evaluate the surrogate is proportional to the product of the number of spatial mesh nodes, $M$, and the cost of a single evaluation of the surrogate, which is of the form $P^{\,\delta}$ where $\delta>0$ depends on the specific interpolation method---i.e., the choice of basis functions as well as the evaluation algorithm.  
It follows that $W^e\simeq MP^{\,\delta}$. 
Recalling the asymptotic estimates for $M$ and $P$ in \eqref{eq:SLSGC_SL_SpatialGridsNo_n_SparseGridsNo}, this implies that the total expense of collecting $N$ surrogate-based Monte Carlo samples is given by 
\begin{equation}
\label{eq:MC_Cost_Surrog_Simu}
\mathcal{W}_\text{SL-SL}^\text{on} = NW^e 
\simeq \epsilon^{-2-\frac{1}{\alpha}-\frac{\delta}{\nu}},
\end{equation}
where the subscript ``SL-SL'' represents the use of single-level stochastic collocation and single-level spatial discretization surrogate for online Monte Carlo sampling.

\begin{remark}[Efficiency of multi-level sampling]
\label{rmk:Efficiency}
The work per sample when using a direct solver for the discrete problem \eqref{eq:FreeBoundary} 
was estimated in \cite{ElLiSa:2022} to be 
proportional to $\epsilon^{-2-\frac{\gamma}{\alpha}}$. 
Hence, whenever the inequality
\begin{equation}
\label{eq:EfficiencyCondition}
\gamma > 1 + \frac{\delta\alpha}{\nu}\,
\end{equation}
is satisfied, the computational complexity of the surrogate-based sampling will be lower than that 
of the direct approach. 
We make a few observations about this relation.
Recall that  $\gamma$ measures the complexity of the linear solver and $\delta$ is a measure of the complexity for the evaluation of the interpolation basis. 
In the particular case of polynomial interpolation with common basis choices such as Chebyshev or Bernstein polynomials, several algorithms for fast evaluation that grow linearly with the degree of the polynomial---such as Horner's \cite{HoGi1819} or Clenshaw's \cite{Clenshaw1955} and multivariate extensions \cite{Carnicer1990,DePe2003,DePe2007}. 
Thus, for high-order polynomial interpolation, the work involved in a single evaluation will grow linearly
with $P$. 
In our setting (high-order Cleshaw-Curtis nodes) it is then reasonable to assume that $\delta=1$. Hence, the computational gain hinges on the ratio between discretization and interpolation rates, which should then satisfy $\gamma-1>\alpha/\nu$ to guarantee an increase in efficiency. Note that for low-order polynomial interpolation the cost can be essentially considered a constant, so that $\delta=0$. For standard sparse direct linear solvers, $\gamma \ge 3/2$ (see further discussion of this
 in Section \ref{sec:Num-Exp}), so that surrogate-based evaluations are always (asymptotically) more efficient than direct solutions. However, if \eqref{eq:EfficiencyCondition} does not hold, the sampling cost of the surrogate model 
may equal or surpass that of the direct solver, particularly as the tolerance $\epsilon$ becomes sufficiently small. Indeed, we have found experimentally that $\gamma\approx 1$ in our tests.
However, if the proportionality constant in the asymptotic expression of the work per sample is smaller for surrogate evaluations, this would still enable surrogate-based sampling to have lower costs; this is what we found in the experiments described in Section \ref{sec:Num-Exp}.
\end{remark}
\subsection{Sampling with surrogates with multilevel spatial discretization}
\label{eq:MLMC-FE-SLML}

We now focus on quantifying the work associated with sampling with the surrogate introduced in Section \ref{sec: SLSGC-ML} that makes use of multiple levels of spatial discretization and a fixed level of accuracy in parameter space, so that in this section $\widehat u_L$ will follow the definition in \eqref{eq:SLSGC-ML_surrogate}.

We express the surrogate at the finest spatial grid level $L$ as a telescoping sum of surrogates on coarser grids, and then exploit the linearity of the expectation to approximate $\mathbb E[u_h]$ as
\[
\mathbb E[u_h] \approx \mathbb{E}\left[\widehat u_{L}\right]=\sum_{\ell=0}^L\mathbb{E}[\widehat{u}_\ell-\widehat{u}_{\ell-1}] =\sum_{\ell=0}^L\mathbb{E}\left[\widehat{Y}_\ell\right],
\]
where the correction term $Y_\ell$ is defined as in \eqref{eq:SLSGC-ML_surrogate}. The expectations $\mathbb{E}[\widehat{Y}_\ell]$ will then be approximated through a Monte Carlo estimator using $N_\ell$ samples at the $\ell$-th spatial mesh level. The resulting unbiased estimator $A[\widehat u_{L}]$  for $\mathbb{E}[u]$ is defined as
\begin{equation}
    \label{eq:MLMC_estimator}
    A\left[\widehat u_{L}\right] := \sum_{\ell=0}^L\frac{1}{N_\ell}\sum_{i=1}^{N_\ell} \widehat{Y}_\ell^{(i)},
\end{equation}
and satisfies ${\displaystyle \mathbb{E}\left[A[\widehat u_h]\right] = \mathbb{E}[\widehat u_{L}]}$ and ${\displaystyle \mathbb{V}\left[A[\widehat u_h]\right] = \sum_{\ell=0}^L \frac{\mathbb{V}\left[\widehat{Y}_\ell^{(i)}\right]}{N_\ell}}$. We remark that for the latter property to hold, the samples $\left(\widehat{u}_\ell^{(i)} - \widehat{u}_{\ell-1}^{(i)}\right)$ comprising the term $\widehat{Y}_\ell^{(i)}$ for each level $\ell$ are drawn \textit{independently of each other} (i.e. the samples at any given level are not reused for the subsequent levels).  However, \textit{within each term} $\widehat{Y}_\ell^{(i)}$, in view of the identity
\[
\mathbb V\left[\widehat{Y}_\ell^{(i)}\right] = \mathbb V\left[\widehat{u}_\ell^{(i)} - \widehat{u}_{\ell-1}^{(i)}\right] = \mathbb V\left[\widehat{u}_\ell^{(i)}\right] + \mathbb V\left[\widehat{u}_{\ell-1}^{(i)}\right] - 2\text{Cov}\left[\widehat{u}_\ell^{(i)},\widehat{u}_{\ell-1}^{(i)}\right]\,,
\]
the strong correlation between $\widehat{u}_{\ell}^{(i)}$ and $\widehat{u}_{\ell-1}^{(i)}$ as $\ell$ grows leads to a reduction in variance $\mathbb{V}\left[\widehat{Y}_\ell^{(i)}\right]$. This contributes to a decrease in the sample size as the mesh gets finer, improving efficiency.

As in the previous section, let $\mathcal{E}^2_{A}$ represent the mean squared error for the multilevel surrogate estimator which can be bounded as
\begin{align}
\nonumber
\mathcal{E}_{A}^2 &=\mathbb E \left[ \left\Vert\mathbb{E}[u]-\mathbb{E}\left[\widehat u_{L}\right]\right\Vert^2_Z 
 \right]+ \mathbb E \left[ \left\Vert \mathbb{E}\left[\widehat u_{L}\right] - A\left[\widehat u_{L}\right] \right\Vert_Z^2\right]=  \underbrace{\left\Vert\mathbb{E}[u]-\mathbb{E}\left[\widehat u_{L}\right]\right\Vert^2_Z}_{\mathcal{E}_{\text{Bias}}^2} \; + \; \underbrace{\mathbb V\left[A\left[\widehat u_{L}\right] \right]}_{\mathcal{E}_{\text{Stat}}^2}\\[1ex]
 \label{eq:StatErr}
 &=\left\Vert\mathbb{E}[u]-\mathbb{E}\left[\widehat u_{L}\right]\right\Vert^2_Z + \sum_{\ell = 0}^L\frac{\mathbb{V}\left[\widehat Y_{\ell}\right]}{N_\ell} \le \underbrace{\bigg(\left\Vert\mathbb E [u] - \mathbb E [u_L]\right\Vert_Z + \left\Vert\mathbb E [u_L]-\mathbb E \left[\widehat u_{L}\right]\right\Vert_Z\bigg)^2}_{\left(\mathcal{E} _{\text{Dis}}+ \mathcal{E}_{\text{Interp}}\right)^2} \; + \; \underbrace{\sum_{\ell = 0}^L\frac{\mathbb{V}\left[\widehat Y_{\ell}\right]}{N_\ell}}_{\mathcal{E}_{\text{Stat}}^2}.
\end{align}
Hence, to achieve a normalized mean squared error below a given tolerance $\epsilon^2$, the split relative discretization, interpolation, and statistical errors should satisfy
\begin{equation}
\label{eq:MLMC_nMSE_ErrSplit}
\overline{\mathcal{E}}_{\text{Dis}} \le \frac{C_m M_L^{-\alpha}}{\left\Vert\mathbb{E}[u_L] \right\Vert_{Z}}\le \theta_1\epsilon, \qquad 
\overline{\mathcal{E}}_{\text{Interp}} \le \frac{C_{p}P^{-\nu}}{\left\Vert\mathbb{E}[u_L] \right\Vert_{Z}}\le \theta_2 \epsilon,\qquad 
\overline{\mathcal{E}}_{\text{Stat}}^2 = \sum_{\ell=0}^L\frac{\mathbb{V}\left[\widehat Y_{\ell}\right]}{N_\ell\left\Vert\mathbb{E}[u_L] \right\Vert_{Z}^2} \le \theta \epsilon^2,
\end{equation}
where, as before, the splitting parameters $\theta, \theta_1,\theta_2\in(0,1)$ are such that $(\theta_1 + \theta_2)^2+\theta = 1$.

The work spent in sampling at one discretization level is proportional to the product of the number of samples collected, $N_\ell$, and the average work of evaluating the surrogate, $W_\ell^e$, both on level $\ell$. The total work is then estimated by adding across levels as
\begin{equation} \label{eq:sampling-cost-SLML} 
\mathcal{W}_{\text{SL-ML}}^\text{on} = \sum_{\ell=0}^L N_\ell W_{\ell}^e,
\end{equation} 
where the superscript ``on" stands for the online work, and the subscript ``SL-ML" denotes the use of a single level in the parameter space and multiple levels in the discretization space. The goal now is to determine the number $N_\ell$ of samples required at every discretization level that will minimize the work requirements while simultaneously yielding a Monte Carlo estimation that satisfies the accuracy threshold $\overline{\mathcal{E}}_{\text{Stat}} \le \sqrt{\theta} \epsilon$. As shown in \cite{ElLiSa:2023,Gi:2008}, this constrained optimization problem yields an estimate for the sample size $N_\ell$  given by
\begin{equation}
\label{eq:MLMC_SampleSize}
N_{\ell} =  \left\lceil \frac{1}{\theta \epsilon^2\left\Vert\mathbb{E}[u_L] \right\Vert_{Z}^2} \sqrt{\frac{\mathbb V\left[\widehat Y_\ell\right]}{W_\ell^e}}\sum_{k = 0}^{L}\sqrt{\mathbb V\left[\widehat Y_k\right] W_k^e}\right\rceil.
\end{equation}
There is, however, one issue with this estimate. For any given level $\ell$, the expression above requires knowledge of the variance $\mathbb{V}\left[\widehat Y_{k}\right]$ for \textit{all levels} $0\leq k\leq L$. In turn, computing or approximating the variance, for instance through the sample variance
\begin{equation}\label{eq:MeanVarUpdate_Var}
\mathbb{V}\left[\widehat Y_{\ell}\right] = \frac{1}{N_\ell-1}\left(\sum_{i=1}^{N_\ell}\left\Vert \widehat Y_{\ell}^{(i)}\right\Vert_{Z}^2-\frac{1}{N_\ell}\left\Vert\sum_{i=1}^{N_\ell}\widehat Y_{\ell}^{(i)}\right\Vert_{Z}^2\right)\,,
\end{equation}
requires knowledge of the sample size $N_\ell$. We address this using a variance extrapolation formula.
Observe that 
\begin{equation*}
    \mathbb{V}[u-\widehat u_\ell] = \mathbb{E}\left[\left\Vert u- \widehat u_\ell\right\Vert_Z^2\right] - \left\Vert \mathbb{E}\left[u- \widehat u_\ell\right]\right\Vert_Z^2 \le \mathbb{E}\left[\left\Vert u-\widehat u_\ell\right\Vert_Z^2\right] =\| u-\widehat u_L\|^2_{L^2(\boldsymbol W, Z)}.
\end{equation*}

As shown in \eqref{eq:SLSGC_ML_Err_exact_surrog}, by selecting the number of sparse grid nodes according to \eqref{eq:SLSGC_ML_SparseGridsNo} and from the bounds (\ref{eq:error-budget}) and (\ref{eq:error-surrogate}),
the total error $\| u-\widehat u_L\|_{L^2(\boldsymbol W, Z)}$ decays at the same rate $\alpha$ as the discretization error, which serves as an upper bound for $\mathbb{V}[u-\widehat u_\ell]$. By virtue of the subtraction in the equation above, this leads to
\begin{equation} \label{eq:decay-rate}
\mathbb{V}[u-\widehat u_\ell] \simeq M_\ell^{-\beta_1},
\end{equation}
where $\beta_1\ge 2\alpha$. Let us also assume that the variance of the difference between surrogates at two successive levels satisfies a similar bound, i.e. $\mathbb{V}[\widehat u_{\ell+1}-\widehat u_\ell] \simeq M_\ell^{-\beta_1}$. With  $\mathbb{V}[\widehat Y_{\ell+1}]=\mathbb{V}[\widehat u_{\ell+1}-\widehat u_\ell]$, and using the mesh growth assumption \eqref{eq:MeshGrowth}, it then follows that
\begin{equation}
\label{eq:VarExtrapolate}
\mathbb V\left[\widehat Y_{\ell+1}\right] \propto (M_\ell/M_{\ell-1})^{-\beta_1}\,\mathbb V\left[\widehat Y_{\ell}\right] = s^{-\beta_1} \mathbb V\left[\widehat Y_{\ell}\right].
\end{equation}
This extrapolation estimate allows us to approximate the variance of subsequent levels based on the sample variances of available levels, giving rise to the following iterative method to estimate the number of required samples at each level. Starting with an initial guess $N_0$ for the number of samples at the coarsest level, the sample variance $\mathbb V[\widehat Y_0]$ is computed, and then for any subsequent level $\ell$  we extrapolate $\mathbb V[\widehat Y_{\ell+1}]$ from $\mathbb V[\widehat Y_{\ell}]$ using \eqref{eq:VarExtrapolate}. The extrapolated values are then substituted into \eqref{eq:MLMC_SampleSize} and the prescribed number of samples is gathered for every discretization level. With these samples the sample variances are computed and a measure of the statistical error, as implicitly defined in the final term of \eqref{eq:StatErr}, is obtained. If the approximation condition \eqref{eq:MLMC_nMSE_ErrSplit} for the statistical error is satisfied, the algorithm stops. Otherwise the number of samples is recomputed using the updated values of the variance and the process is repeated until condition \eqref{eq:MLMC_nMSE_ErrSplit} is met. The process is described in Algorithm \ref{algo:MLMC_Algo_CorrectionVersion}. This algorithm is derived from Giles' basic theorem \cite{Gi:2008} and it accounts for both discretization and statistical errors.  A key difference is that the algorithm uses a surrogate model to evaluate samples, with the interpolation error reflected through the construction of a surrogate in the algorithm input.

\normalem
\begin{algorithm}[!ht]
\label{algo:MLMC_Algo_CorrectionVersion}
\DontPrintSemicolon
   
   \KwIn{Initial spatial discretization mesh 
    $\mathcal{T}_0$, initial sample size 
    $\boldsymbol N_{\text{old}}$, 
        tolerance $\epsilon$, splitting parameter $\theta\in (0,1)$, single-spatial grid 
        surrogate $\widehat{u}_0$ constructed with $P$ sparse grid nodes.}\vspace{1ex}
    
    \KwOut{Fine-grid level $L$, vector of sample sizes 
    $\boldsymbol N = (N_0\ldots,N_L)$ for each discretization level, vector of variance estimations 
    $\boldsymbol V =(\mathbb V[\widehat Y_0],\ldots,\mathbb V[\widehat Y_L])$, expectation 
    estimate $A$.}\vspace{1ex}
    \hrule \vspace{1ex}
    
    Initialize level $L \leftarrow 0$.
   
      Initialize vector of sample number corrections $\boldsymbol{dN} \leftarrow \boldsymbol 
         N_{\text{old}}$   
         
     \While{$\sum_\ell dN_\ell>0$\,}{
    \For{$0\le \ell\le L$\,}{
    
        \For{$i = 1,\ldots,dN_\ell $\,}
    {
    Evaluate the surrogate on $\mathcal{T}_\ell$ to obtain $\widehat{u}_{\ell}^{(i)}$.
    }
    }
     Approximate $\mathbb V[\widehat Y_{\ell}]$ for $0\leq\ell\leq L$ using \eqref{eq:MeanVarUpdate_Var}.
        
     Update the sample size estimation $N_\ell$  for $0\leq\ell\leq L$ by \eqref{eq:MLMC_SampleSize}.  
    
    $\boldsymbol{dN} \leftarrow \max(0,\boldsymbol N-\boldsymbol N_{\text{old}})$. 
    
    $\boldsymbol N_{\text{old}}\leftarrow \boldsymbol N$.
    
    \If{$\sum_\ell dN_\ell=0$}{
    
    \If{ $\left\Vert u - u_L\right\Vert_{L^2(\boldsymbol W,Z)} < \theta_1 \epsilon \left\Vert\mathbb{E}[u] \right\Vert_{Z}$,}
    {
    Compute $A[\widehat u_L]$ by \eqref{eq:MLMC_estimator}. 
    }
    \Else {
     $L \leftarrow L+1$.
     
     Approximate $\mathbb V[\widehat Y_L]$ by  \eqref{eq:VarExtrapolate}.
     
     Increase the length of the vector $\boldsymbol{N}$ by appending the value $N_L$ prescribed by \eqref{eq:MLMC_SampleSize}.

     Increase length of the vector $\boldsymbol{N}_{\text{old}}$ by appending $0$.
     
     $\boldsymbol{dN} \leftarrow \boldsymbol N-\boldsymbol N_{\text{old}}.$
     }
    }
    }  
\caption{Multi-level Monte Carlo Finite-Element}
\end{algorithm}
\ULforem

We now return to the quantification of the computational work required for sampling. 
Using the value for $N_\ell$ given by \eqref{eq:MLMC_SampleSize}, it follows that the online sampling cost
\[
\sum_{\ell=0}^L N_\ell W_{\ell}^e \geq \sum_{\ell=0}^L \left(\frac{1}{\theta \epsilon^2\left\Vert\mathbb{E}[u_L] \right\Vert_{Z}^2} \sqrt{\frac{\mathbb V\left[\widehat Y_\ell\right]}{W_\ell^e}}\sum_{k = 0}^{L}\sqrt{\mathbb V\left[\widehat Y_k\right] W_k^e}\right)W_{\ell}^e = \frac{1}{\theta \epsilon^2 \left\Vert\mathbb{E}[u_L] \right\Vert_{Z}^2}\left(\sum_{\ell=0}^L \sqrt{\mathbb V\left[\widehat Y_\ell\right] W_\ell^e}\right)^2.
\]
Underestimating the sum on the right using only the final term $\ell=L$ yields
\begin{equation} \label{eq:underest-sum}
\sum_{\ell=0}^L N_\ell W_{\ell}^e \geq 
\frac{\mathbb V\left[\widehat Y_L\right] W_L^e}{\theta \epsilon^2 \left\Vert\mathbb{E}[u_L] \right\Vert_{Z}^2},
\end{equation}
and recalling that the work to evaluate the surrogate is $W_L^e \simeq M_LP^{\,\delta}$ (see the discussion following (\ref{eq:MC_SampleSize})) with $M_L\simeq \epsilon^{-1/\alpha}$ and $P\simeq \epsilon^{-1/\nu}$, we obtain 
\begin{equation} \label{eq:sampling-work-lower}
W_L^e \simeq \epsilon^{-1/\alpha} \epsilon^{-\delta/\nu} = \epsilon^{-\frac{1}{\alpha}-\frac{\delta}{\nu}}.
\end{equation}
Moreover, (\ref{eq:VarExtrapolate}) implies that
\begin{equation}\label{eq:VarianceExpansion}
\mathbb V\left[\widehat Y_L\right] 
= s^{-\beta_1 L} \mathbb V\left[\widehat Y_0\right]
= s^{-\beta_1 L} \mathbb V\left[\widehat u_0\right],
\end{equation}
while and \eqref{eq:SLSGC_MLS_SpatialGridsNo} implies
\begin{equation}\label{eq:Lepsilon}
L\simeq \frac{1}{\alpha} \log_s \left(\epsilon^{-1}\right).
\end{equation}
These lead to 
\[
\mathbb V\left[\widehat Y_L\right] = s^{-\beta_1 L} \mathbb V\left[\widehat u_0\right] 
\simeq s^{-\frac{\beta_1}{\alpha}\log_s\left(\epsilon^{-1}\right)} \simeq \epsilon^{\frac{\beta_1}{\alpha}}.
\]
Combining this with (\ref{eq:underest-sum}) and (\ref{eq:sampling-work-lower}) gives the lower bound
\[
\sum_{\ell=0}^L N_\ell W_{\ell}^e \geq c(\epsilon), \quad \mbox{where} \quad 
c(\epsilon) \simeq
\left(\frac{1}{\theta \epsilon^2 \left\Vert\mathbb{E}[u_L]\right\Vert_{Z}^2} \right) 
\epsilon^{2} \epsilon^{-\frac{1}{\alpha}-\frac{\delta}{\nu}} \simeq 
\epsilon^{-\frac{1}{\alpha}-\frac{\delta}{\nu}}.
\]
For an upper bound, we start by noting that $\lceil x\rceil\leq 2x$ for all $x\geq 0$ and thus, from \eqref{eq:MLMC_SampleSize} we have
\[
\sum_{\ell=0}^L N_{\ell}W_\ell^e \leq  \sum_{\ell=0}^L\left(\frac{2}{\theta \epsilon^2\left\Vert\mathbb{E}[u_L] \right\Vert_{Z}^2} \sqrt{\frac{\mathbb V\left[\widehat Y_\ell\right]}{W_\ell^e}}\sum_{k = 0}^{L}\sqrt{\mathbb V\left[\widehat Y_k\right] W_k^e}\right)W_\ell^e = \frac{2}{\theta \epsilon^2\left\Vert\mathbb{E}[u_L] \right\Vert_{Z}^2}\left(\sum_{\ell=0}^L\sqrt{\mathbb V\left[\widehat Y_k\right] W_k^e}\,\right)^2.
\]
We now focus our attention on the term $\sum_{k = 0}^{L}\sqrt{\mathbb V\left[\widehat Y_k\right] W_k^e}$ and note that from \eqref{eq:MeshGrowth} and \eqref{eq:SLSGC_ML_SparseGridsNo} it follows that
\begin{equation}\label{eq:Wk}
W_k^e = M_kP^\delta = M_k M_L^{\alpha\delta/\nu} = M_0^{1+\alpha\delta/\nu}s^{k+ L\alpha\delta/\nu} \simeq s^{k+ L\alpha\delta/\nu}.
\end{equation}
Therefore, using \eqref{eq:VarianceExpansion} and \eqref{eq:Wk} first, and then \eqref{eq:Lepsilon} we obtain
\[
\sum_{k = 0}^{L}\sqrt{\mathbb V\left[\widehat Y_k\right] W_k^e} \simeq s^{L\alpha\delta/2\nu}\sum_{k = 0}^{L}s^{k(1-\beta_1)/2} \simeq \epsilon^{-\delta/2\nu}\sum_{k = 0}^{L}s^{k(1-\beta_1)/2}.
\]
Depending on the value of $\beta_1$, Lemma \ref{lem:lem1} below prescribes the value of the sum on the right hand side of the expression above. From this we can now conclude that
\[
\sum_{\ell=0}^L N_{\ell}W_\ell^e \lesssim \begin{cases} \epsilon^{-2-\frac{\delta}{\nu}} & \text{ if }\;\beta_1 >1, \\[1ex] 
\epsilon^{-2-\frac{\delta}{\nu}}\, |\log \epsilon|^2 & \text{ if }\;\beta_1 = 1, \\[1ex]
\epsilon^{-2-\frac{\delta}{\nu}\,-\,\frac{1-\beta_1}{\alpha}} & \text{ if }\;\beta_1 < 1 .  \end{cases}
\]
Thus, in light of (\ref{eq:sampling-cost-SLML}), we have established the following bounds on sampling 
cost with multilevel spatial discretization:
\begin{theorem}
\label{thm:SLSGC-ML_sampling_cost}
The costs of sampling using a surrogate derived from multilevel spatial discretization are bounded
above and below as
\[
\epsilon^{-\frac{1}{\alpha}-\frac{\delta}{\nu}} \lesssim \mathcal{W}_{\text{\rm SL-ML}}^\text{\rm on} \lesssim
\begin{cases} \epsilon^{-2-\frac{\delta}{\nu}} & \text{ if }\;\beta_1 >1, \\[1ex] 
\epsilon^{-2-\frac{\delta}{\nu}}\, |\log \epsilon|^2 & \text{ if }\;\beta_1 = 1, \\[1ex]
\epsilon^{-2-\frac{\delta}{\nu}\,-\,\frac{1-\beta_1}{\alpha}} & \text{ if }\;\beta_1 < 1.
\end{cases}
\]
\end{theorem}
The theorem states that if $\beta_1>1$ (as is the case in the experimental tests described in the next section), then the primary sampling occurs on the coarsest grid. 
For smaller values of $\beta_1$, fine grids will play a larger role in costs, and the benefits of multilevel spatial discretization are limited.\footnote{This discussion allows the convergence rate of the variance to be larger than $2\alpha$ in \eqref{eq:decay-rate} and \eqref{eq:VarExtrapolate}. The worst-case scenario is $\beta_1=2\alpha$. With $\alpha$ as specified in \eqref{eq:error-budget}, $\alpha=1$ for smooth two-dimensional problems and $\alpha=2/3$ for smooth three-dimensional problems, giving significant reductions of fine-grid sampling costs in these scenarios.}

\begin{lemma}\label{lem:lem1}
Let $\eta\in \mathbb{R}$. The behavior of $\,\sum_{\ell=0}^L s^{\eta \ell}$ is described by
\[
\sum_{k=0}^L s^{\eta k} \simeq \left\{\begin{array}{ll}
\frac{1}{1-s^\eta}, & \eta<0,\\
|\log \epsilon|, & \eta = 0,\\
\epsilon^{-\eta/\alpha}, & \eta>0.
\end{array}
\right.
\]
\end{lemma}

\section{Numerical experiments}\label{sec:Num-Exp}

We now present numerical results for surrogate-based sampling to estimate the expected poloidal flux $\mathbb E[u]$. We build surrogates through sparse grid collocation and explore the construction cost. We then perform MC-FE and MLMC-FE sampling with both direct computation and surrogates, assessing efficiency by comparing the computational cost, as measured by CPU times, and also evaluating the accuracy of certain quantities such as the separatrix and various geometric descriptors derived from the approximation of $\mathbb E[u]$. We note at the outset that in previous work \cite{ElLiSa:2022}, 
we found that the interpolation error appears to satisfy a bound like (\ref{eq:coll-error-bound-2}) but not (\ref{eq:coll-error-bound-1}), so that the cost analysis of the previous sections is not directly applicable. Despite this, the results suggest that collocation and multilevel MC lead to cost savings along the lines suggested by the analysis.

The coils are modeled as independent and uncorrelated random variables, each following a uniform distribution centered around a baseline value $I_k$ and subject to a perturbation of relative size $\tau=2\%$. The joint density function $\pi(\boldsymbol{\omega})$ and the $d$-dimensional parameter space $\boldsymbol W$ are then given by
\begin{equation*}
\label{eq:ParameterSpace}
 \pi \left(\boldsymbol{\omega}\right)=\prod_{k=1}^{d} \pi_k\left(\omega_{k}\right)=\prod_{k=1}^{d} \frac{1}{2\tau |I_k|}, \qquad  
   \boldsymbol W := \prod_{k=1}^{d}\left[I_k-\tau \left\vert I_k\right\vert,I_k+\tau \left\vert I_k \right\vert\right],
\end{equation*}
where we use the values of \cite{FaHe:2017} as the baseline values of current intensities, giving a vector $\boldsymbol{I}$ of length twelve:
\[
{\renewcommand{\arraycolsep}{2pt}
\begin{array}{lllllll}
I_1 = -1.4 \times 10^{6} A, & \phantom{+} & I_2 = -9.5 \times 10^{6} A, & \phantom{+} & I_3 = -2.0388 \times 10^{7} A, & \phantom{+} & I_4 = -2.0388 \times 10^{7} A, \\
I_5 = -9 \times 10^{6} A, & \phantom{+} & I_6 = 3.564 \times 10^{6} A, & \phantom{+} & I_7 = 5.469 \times 10^{6} A, & \phantom{+} & I_8 = -2.266 \times 10^{6} A, \\
I_9 = -6.426 \times 10^{6} A, & \phantom{+} & I_{10} = -4.82 \times 10^{6} A, & \phantom{+} & I_{11} = -7.504 \times 10^{6} A, & \phantom{+} & I_{12} = 1.724 \times 10^{7} A. 
\end{array}
}
\]
For every realization of $\boldsymbol \omega$, we made use of the finite element-based solver {\tt FEEQS.m} \cite{FaHe:2017,Heumann:feeqsm,CEDRES}, developed by H.\ Heumann and collaborators to obtain a numerical solution of \eqref{eq:FreeBoundary}. When solving the  perturbed problem, the solution corresponding to the baseline values for the currents presented above was used as initial guess for Newton's iterations.

\subsection{Experiment description}
MLMC-FE sampling requires a hierarchical set of spatial meshes. 
Starting from a \textit{reference mesh} provided by {\tt FEEQS.M},
we generate this set using the {\tt Triangle} mesh generator \cite{Sh:2002} to refine and derefine the reference mesh within the region enclosed by the (spline-approximated) boundaries. This process results in a collection of non-nested and 
geometry-conforming uniform meshes \cite{ElLiSa:2023}. The number of grid points $M_\ell$ of each mesh in the hierarchy is detailed in Table \ref{Tab:Dof}, where the reference mesh (level 2 in the hierarchy) contains 30,449 grid points. We generate surrogates on these spatial meshes using stochastic collocation on Clenshaw-Curtis sparse grids within a 12-dimensional parameter space. The number of sparse grid nodes $P_q$ at increasing levels is shown in Table \ref{Tab:Dof}. 
For all sampling conducted in this study, the tolerances for the normalized mean squared error range from $\epsilon=2\times 10^{-4}$ to $8\times 10^{-3}$. The sample mean and the normalized sample variance $V_{\ell} \equiv \mathbb V[\widehat Y_\ell]/\left\Vert\mathbb{E}[u_L] \right\Vert_{Z}^2$ are dynamically updated using Welford’s algorithm
\begin{table}[ht]
\centering
\scalebox{0.8}{
\begin{tabular}{c|c|c|c|c|c|c|c|c|c|c|c|c|c|c|c|c|c|c|}
\cline{1-7}	
\multicolumn{1}{|c|}{ Level} &0&1&2&3&4&5\\
\hline
\multicolumn{1}{|c|}{$M_\ell$}&2685&8019&30449&120697&484080&1934365\\
\hline
\multicolumn{1}{|c|}{$P_q$}&1 &25 &313 &2649 &17265&--\\
\hline
\end{tabular}}
\caption{Number of spatial grid points $M_\ell$ at increasing spatial grid level $\ell = 0$ to 5, and number of sparse grid points $P_q$ from sparse grid level $q = 0$ to 4. In the numerical experiments reported in this paper, we use all spatial levels $\ell = 0,\ldots,5$, while the sparse grid level is fixed to $q=1$ due to computational limitations.}
\label{Tab:Dof}
\end{table}
In previous work \cite{ElLiSa:2023}, we showed that the L2-norm of the discretization error decays at a rate of 
approximately one ($\alpha\approx1$), as expected for piecewise linear elements. We also showed in \cite{ElLiSa:2022} that the interpolation error decays slowly for small $P$, suggesting that the error behaves like (\ref{eq:coll-error-bound-2}). Limited computational resources make it difficult to use a large enough level $q$ for collocation to satisfy (\ref{eq:coll-error-bound-2}) for small $\epsilon$, and in these tests we limited our choice to level $q=1$.

For direct non-linear solves at each parameter value, the stopping threshold for the relative residual of Newton's method is set to $5\times 10^{-11}$. The normalization factor $\|\mathbb{E}[u_\ell]\|_Z$ on the finest spatial mesh level ($\ell=5$) is approximately $8.5708\times 10^{-1}$. All experiments are conducted using {\tt MATLAB} R2023a on a System 76 Thelio Major with 256GB RAM and a 64-Core @4.6 GHz AMD Threadripper 3 processor. All experiments are conducted using {\tt MATLAB} R2023a on a System 76 Thelio Major with 256GB RAM and a 64-Core @4.6 GHz AMD Threadripper 3 processor.
%
\subsection{Surrogate construction (offline) costs}
%
In this section, we briefly discuss aspects of the cost of constructing surrogates.
We start with Table \ref{Tab:Finest_Level_Surrg}, which shows the spatial grid levels (see (\ref{eq:SLSGC_MLS_SpatialGridsNo})) required to make an estimate of the discretization error less than a variety of tolerances $\theta_1\epsilon$ for $\theta_1=\sqrt{0.5}/2$. The results shown are for surrogates built from both single-level spatial methods and multilevel spatial methods, where for both, the discretization error is estimated using an a posteriori error estimator \cite{EiMeNe:2016,KhPaHe:2020}; see \cite{ElLiSa:2023} for details.

\begin{table}[ht]
\centering
\scalebox{0.8}{
\begin{tabular}{c|c|c|c|c|c|c|c|c|c|c|c|c|c|c|c|c|c|c|}
\hline
\multicolumn{1}{|c|}{Tolerance} &$4\times 10^{-4}$&$6\times 10^{-4}$&$8\times 10^{-4}$&$10^{-3}$&$2\times 10^{-3}$&$4\times 10^{-3}$&$6\times 10^{-4}$&$8\times 10^{-3}$\\
\hline
\multicolumn{1}{|c|}{Finest spatial grid level $L$}&5&5&4&4&4&3&3&3\\
\hline
\end{tabular}}
\caption{Finest spatial grid levels required for surrogates to meet various error tolerances.}
\label{Tab:Finest_Level_Surrg}
\end{table}

\begin{figure}[ht!]\centering
\begin{tabular}{cc}
\includegraphics[width=0.46\linewidth]{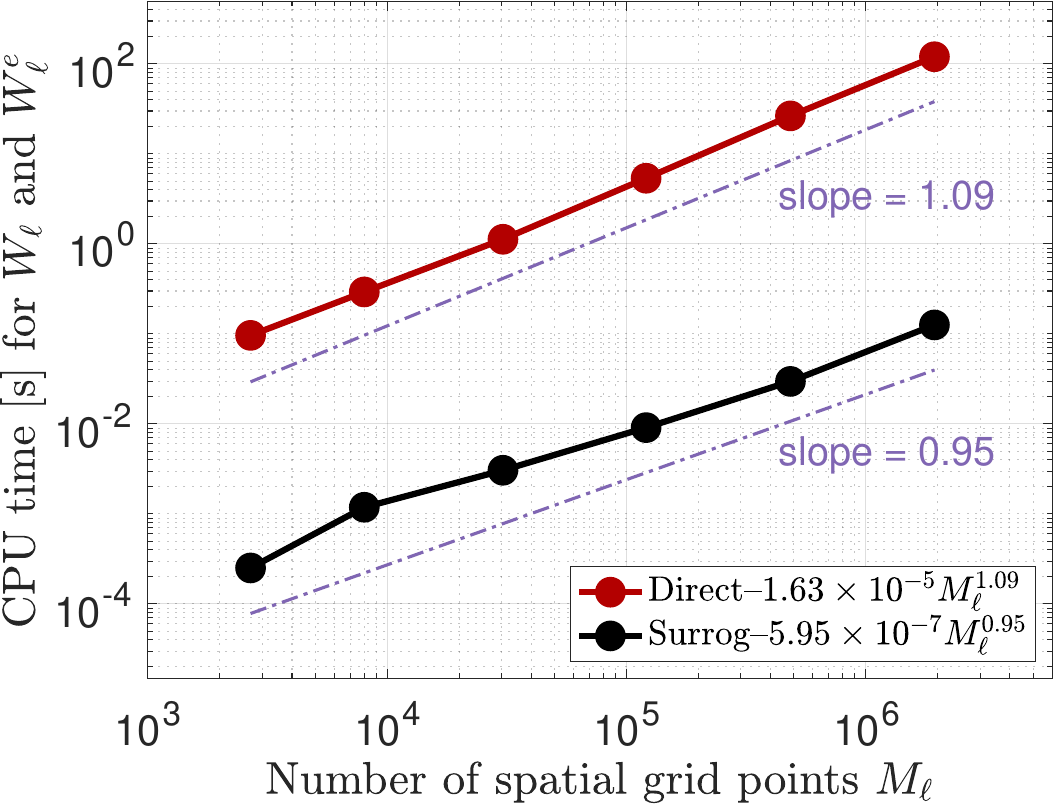}&
\includegraphics[width=0.46\linewidth]{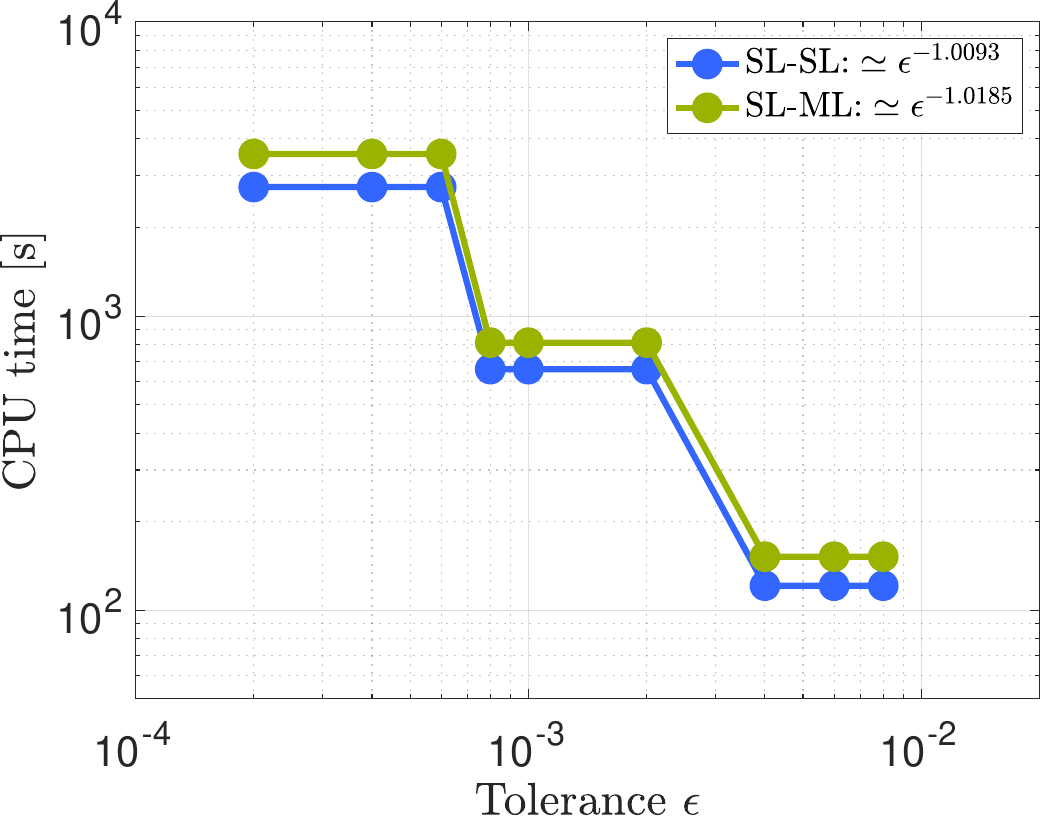}
\end{tabular}
\caption{Left: Mean CPU time per realization vs. number of spatial grid points $M_\ell$ for both direct and surrogate computations. The computing time was obtained by averaging over 100 realizations. Right: Offline costs of construction of both single spatial level and multi spatial level surrogates.}
\label{fig:CostEstimatePlot}
\end{figure}

The cost of constructing the surrogate is then the cost of computing the direct solution on each sparse grid node times the number of sparse grid nodes. For these tests, the latter number is $25$, corresponding to sparse grid level $q=1$. We explored the cost of a single direct solution in \cite{ElLiSa:2023}. These are summarized in the blue curve shown in the right panel of Figure \ref{fig:CostEstimatePlot} (This part of the image is reproduced from \cite{ElLiSa:2023}). They were obtained by solving the systems for 100 random currents and taking the mean CPU time, and they indicate that the costs were of the form $M^\gamma$ for $\gamma=1.09$. The resulting total costs are summarized in Table \ref{Tab:Time_construct_surrogate} and plotted in the right image of Figure \ref{fig:CostEstimatePlot}, for both single-level and multilevel (in space) surrogates. The plots indicate that the two surrogates have asymptotic complexity of magnitude $\epsilon^{-1}$, which aligns with the theoretical cost $\epsilon^{-\gamma/\alpha}\approx \epsilon^{-1.09}$. The slightly larger cost of the multilevel version is due to the fact that it requires the surrogates for all coarser grids, as observed in the comments that follow (\ref{eq:SLSGC_MLS_Cost_Construct}).
\begin{table}[ht]
\centering
\scalebox{0.8}{
\begin{tabular}{c|c|c|c|c|c|c|c|c|c|c|c|c|c|c|c|c|c|c|}
\hline
\multicolumn{1}{|c|}{Spatial grid level $\ell$} &0&1&2&3&4&5\\
\hline
\multicolumn{1}{|c|}{CPU time (SL-SL)\ }&2.02e+00&5.95e+00&2.30e+01&1.21e+02&6.60e+02&2.75e+03\\
\multicolumn{1}{|c|}
{CPU time (SL-ML)}
&2.02e+00&7.90e+00&3.10e+01&1.52e+02&8.12e+02&3.56e+03\\
\hline
\end{tabular}}
\caption{CPU time to construct surrogate $\widehat u_{\ell,q}$ of sparse 
grid level $q=1$ with respect to increasing spatial grid levels. 
The time for SL-SL is calculated using the cost per sample of direct 
computation multiplied by the number of sparse grid nodes of sparse grid level 1. The time for SL-ML at level $\ell$ is the sum of the single-level times for
all levels less than or equal to $\ell$.}
\label{Tab:Time_construct_surrogate}
\end{table}
\subsection{Memory considerations} 
In addition to computational time, it is important to compare the memory requirements of the surrogate construction with those of the direct solver. For the P1 finite element discretization considered here, the stiffness matrices are sparse with a uniformly bounded number of non-zero entries per row under standard mesh regularity assumptions. Consequently, the memory required for a single direct solve at level $\ell$ scales linearly with the number of spatial degrees of freedom $M_\ell$. For surrogate construction, additional memory costs arise from storing the direct solutions at each sparse grid node. If $P_q$ denotes the number of sparse grid nodes and $L$ the finest spatial level required to meet a given tolerance, then the memory required for the single-level spatial surrogate scales as $\mathcal{O}(P_q M_L)$. In the multilevel case, solutions on all levels $\ell=0,\dots,L$ must be retained, yielding a memory requirement of $\mathcal{O}\!\left(P_q \sum_{\ell=0}^{L} M_\ell \right)$. Since $M_\ell$ grows geometrically with $\ell$, this sum is dominated by $M_L$, and thus the multilevel memory cost is also asymptotically $\mathcal{O}(P_q M_L)$.

In the present experiments, the sparse grid level is fixed at $q=1$, so $P_q=25$ remains constant as the tolerance $\epsilon$ decreases. Accordingly, the memory usage increases only through the growth of $M_L$ as finer spatial grids are required (Table~\ref{Tab:Finest_Level_Surrg}). The surrogate therefore introduces a moderate multiplicative memory factor proportional to $P_q$, while preserving the same asymptotic scaling with respect to spatial resolution as the direct approach.

\subsection{Sampling (online) costs}
%

We now use these surrogates to run online simulations with MC-FE and MLMC-FE sampling. For each sampling approach, we will examine the sampling cost, calculate the statistical estimator for the expected solution \eqref{eq:QoI}, estimate the sample size, and compare the combined offline and online costs. Note that it follows from (\ref{eq: Smolyak_Quad_formula}) that the average cost to evaluate the surrogate on a grid with $M_{\ell}$ points is linear in $M_\ell$. It can also be shown that the dependence on the number of sparse grid nodes is proportional to $M_\ell P_q^{\delta}$ \cite[Section 3.3.5.4]{KlBa:2004}. We have found experimentally that $\delta\approx 1$ \cite{Li:2024}.

\begin{figure}[!t]\centering
\begin{tabular}{cc}
\includegraphics[width=0.46\linewidth]{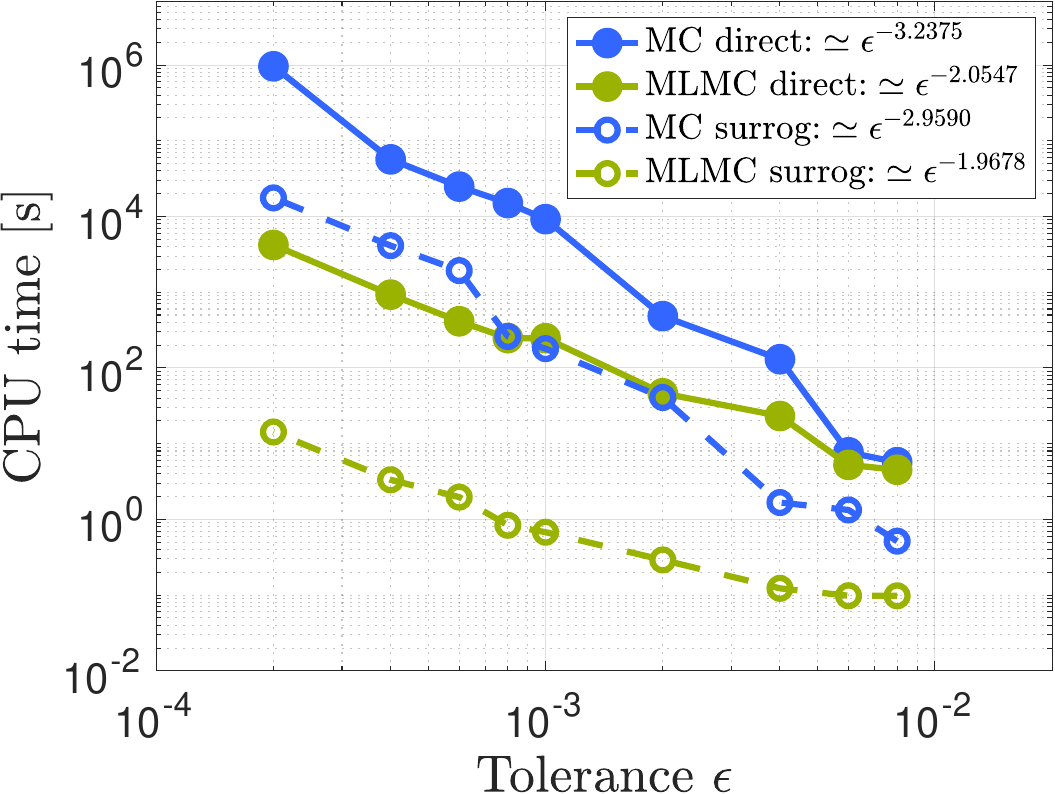}&
\includegraphics[width=0.46\linewidth]{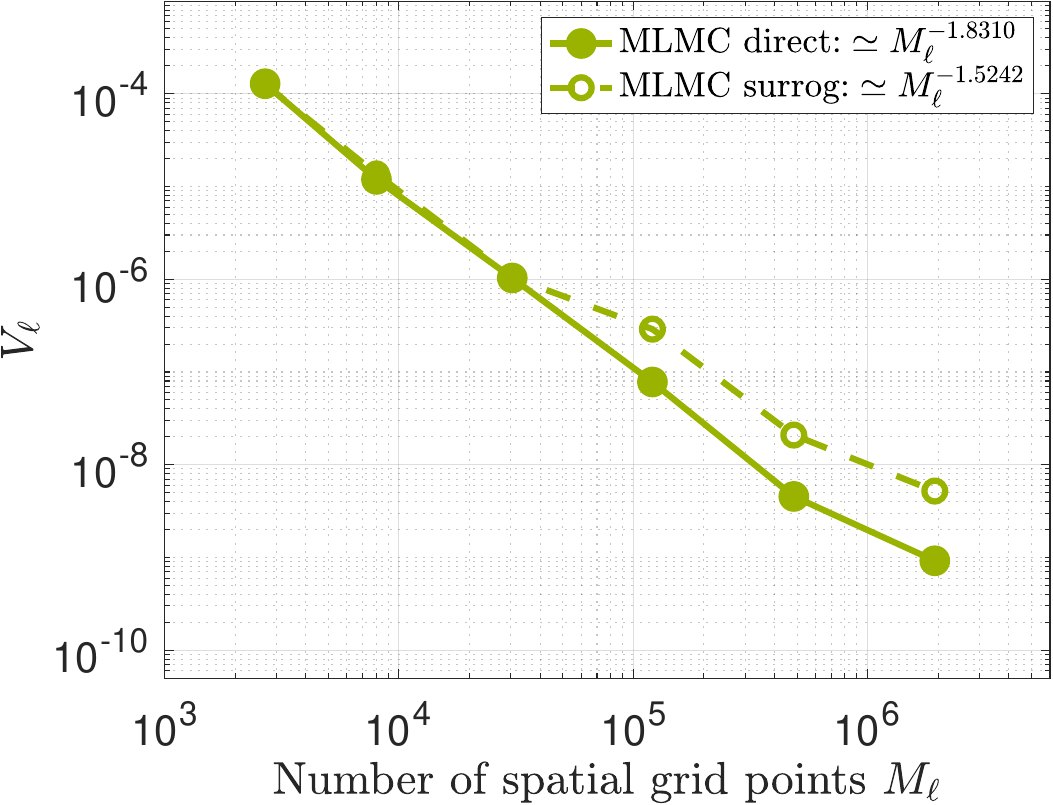} 
\end{tabular}
\caption{
Left: Estimated sampling CPU time vs.\ tolerance $\epsilon$.
Right: value of $V_\ell$ vs.\ number of spatial grid points $M_\ell$. 
} 
\label{fig:CollectionPlot} 
\end{figure}

The online costs in CPU time associated with the two sampling techniques, both for direct solve and surrogate evaluation, are illustrated in the plot on the left of Figure \ref{fig:CollectionPlot}. Since we use a fixed $q=1$ for building the surrogate, the number of sparse grid nodes $P_1$ is independent of $\epsilon$, and the online cost \eqref{eq:MC_Cost_Surrog_Simu} behaves like $\epsilon^{-2-\frac{1}{\alpha}}$. 
Least-squares fits indicate that the direct solve and surrogate evaluation 
exhibit computational costs of magnitudes $\epsilon^{-3.24}$ and $\epsilon^{-2.96}$ respectively, aligning closely with the theoretical values of $\epsilon^{-2-\frac{1}{\alpha}}=\epsilon^{-3}$ discussed in  Remark \ref{rmk:Efficiency} and $\epsilon^{-2-\frac{\gamma}{ \alpha}} = \epsilon ^{-3}$. (Here the term $\epsilon^{-\frac{\delta}{\nu}}$, which comes from the now
fixed number $P_1$ of sparse grid nodes, is incorporated into the constant.)

Since $\gamma \approx 1$, the solid and dotted blue lines in the plot are almost parallel.\footnote{Note that the value $\gamma=1.09$ was determined from experiments.
In general, for large $M$, we expect $\gamma$ to be larger for sparse direct solvers of the Jacobian systems \cite{Da:2006,GoVaCh:2013}, such as $\gamma=1.5$ for the nested dissection method \cite{Ge:1973} and closer to $2$ for a band solver \cite[Section 4.3]{GoVaCh:2013}.}
The cost of the methodology used for direct solution (solid blue curve) is influenced by various factors, including the cost of solving the Jacobian system, the number of iterations required by Newton's method, complexities from the nonlinearity of the free plasma boundary, and the need to reassemble the nonlinear system. We believe that these factors lead to the higher cost of the direct solution indicated in the figure. If $\gamma$ were larger, the advantages of the surrogate solution (dashed blue curve) would be more dramatic.

For the MLMC-FE sampling approaches, the solid and dashed green curves in the left image of Figure \ref{fig:CollectionPlot} indicate that the sampling costs of both the direct solve and surrogate methods follow a complexity proportional to $\epsilon^{-2}$, consistent with the theoretical predictions in Theorem \ref{thm:SLSGC-ML_sampling_cost}. Notably, the surrogate approach, represented by the dashed green line, incurs significantly lower costs than the direct method, shown in the nearly parallel solid green line. This disparity can be attributed to a smaller constant factor in the asymptotic estimate for the sample-wise cost for surrogate evaluation, as evidenced in the left images of Figure \ref{fig:CostEstimatePlot}. The fitted curves in that image indicate that this factor behaves like $1.63\times 10^{-5} \times M_\ell^{1.09}$ for direct evaluation and $3.08\times 10^{-8} \times M_{\ell}^{0.95} P_1^{0.92}$ for surrogate evaluation. Taking the exponents of $M_{\ell}$ and $P_1$ to be $1$ and replacing (fixed) $P_1$ with $25$, these costs simplify to $1.63\times 10^{-5}M_{\ell}$ and $7.7\times 10^{-7}M_{\ell}$. Moreover, in general, for both direct and surrogate methods, the multilevel versions of sampling are less costly than the single-level ones.

The online costs of multilevel methods are influenced not only by the cost of computing a single sample but also by the values of $V_\ell$, which determine the required sample sizes \eqref{eq:MLMC_SampleSize}. The plot on the right of Figure \ref{fig:CollectionPlot} shows $V_\ell$ for both direct and surrogate evaluation. The observed decay rates of $V_\ell$ are approximately $1.83$ for direct computation and $1.52$ for surrogate. For the direct computation, the decay rate in \eqref{eq:decay-rate} of $\beta_1\approx 2\approx 2\alpha$ aligns with the worst-case scenario for $\beta_1$. This result is consistent with findings reported in \cite{ElLiSa:2023}. In contrast,  the surrogate exhibits a slightly lower decay rate, resulting in slightly higher values of $V_\ell$ compared to the direct solver when the sample size $M_\ell$ exceeds $10^5$. This discrepancy arises from an insufficient number of sparse grid nodes to construct the surrogate. Recall that to ensure accuracy, the number of sparse grid nodes must satisfy \eqref{eq:MLMC_nMSE_ErrSplit}. We used a fixed sparse grid level $q=1$, leading to a nearly constant interpolation error, violating the accuracy requirement. Consequently, as the tolerance falls below $10^{-3}$ (when $M_\ell$ exceeds $10^5$), the large interpolation error (around $10^{-3}$ in the plot) becomes more significant than the discretization error. This increases the total error $\|u-\widehat u_\ell\|$ for the surrogate approach, resulting in higher values of $V_\ell$ than for the direct solver. Per Theorem \ref{thm:SLSGC-ML_sampling_cost}, the majority of the work for multilevel methods is concentrated on the coarse spatial grids for both the surrogate-enhanced MLMC-FE and the direct solver counterpart, effectively shifting the workload from the fine spatial mesh to the coarse grids and enhancing computational efficiency.

The MLMC-FE estimator for \eqref{eq:QoI} involves accumulating surrogate sample corrections 
$\widehat{Y}_{\ell}^{(i)}$ across $L$ spatial levels. As we discussed in previous work \cite{ElLiSa:2023}, for each sample correction,  $\widehat{u}_{\ell-1}^{(i)}$ is constructed on the coarse spatial grid by interpolating $\widehat{u}_{\ell}^{(i)}$ from the fine grid.  This interpolation method serves as a more efficient, though slightly less accurate, alternative to the Galerkin projection. However, the use of non-nested geometry-conforming uniform spatial meshes introduces extrapolation errors during interpolation. This issue can be addressed using a sufficiently fine common grid
(e.g., $\ell=5$) that encompasses all coarser meshes, in order to minimize or eliminate extrapolation errors caused by interpolation of surrogate corrections from all multi-level spatial coarser grids. Although this does increase accuracy, it does so at a significant extra cost.

Table \ref{Tab:CPU_time_2} provides a quantitative summary of the online CPU time for the various solution strategies, including interpolation to a common fine grid just discussed. The discrete solution corresponding to the baseline currents presented at the beginning of Section \ref{sec:Num-Exp} was used as the initial guess for all non-linear iterations. Since all samples share the same geometry, mesh, and finite element spaces, the associated data structures and assembly routines are initialized once at the beginning of the computation. Thus, the reported solution times reflect primarily the nonlinear solve and linear system solution for each realization. The one-time setup cost is small compared with the cumulative cost of the many solves required for Monte Carlo or multilevel sampling, and is therefore effectively amortized and negligible in the regimes considered here. The entries in the table correspond to the time required to make the square root of the sum of the squares of the discretization and statistical errors (see (\ref{eq:MC_nMSE_ErrSplit})) less than the tolerances, since the interpolation error is limited by the fixed choice of a sparse grid. The table uses the Monte Carlo method with direct computation as a benchmark and calculates the speedups of the different sampling strategies. It is evident that simply replacing the direct solve with the surrogate 
(third column) produces significant speedups (on the order of 10 to 50); replacing the full grids with multilevel grids (fourth column) results in somewhat better speedups (up to 200 for small tolerance), and using both surrogates and multilevel grids (fifth column) yields dramatic speedups, often exceeding $10^4$. The last two columns of the table present results for mitigating extrapolation errors using a common fine grid for MLMC with both direct solve and surrogate, revealing that the speedups for both approaches are comparable due to the significant interpolation cost. In the following, we discuss some issues regarding accuracy.

\begin{table}[ht]
\centering
\scalebox{0.52}{\large
\begin{tabular}{c|c|c|c|c|c|c|c|c|c|c|c|c|}
\cline{2-9}
 & \multicolumn{4}{|c|}{\large \textbf{Timings without post--processing}} & \multicolumn{4}{|c|}{\large \textbf{ Timings with post--processing}} \\
\hline
\multicolumn{1}{|c|}{ }&  &  &  &  & Interpolation & Interpolation  & Heat flow & Heat flow  \\
\multicolumn{1}{|c|}{ }& MC-FE & MC-FE & MLMC-FE & MLMC-FE & MLMC-FE  &MLMC-FE & MLMC-FE & MLMC-FE \\
\multicolumn{1}{|c|}{\huge$\epsilon$}&Direct solver &Surrogate& Direct solver &Surrogate &Direct solver &Surrogate &Direct solver &Surrogate\\
\multicolumn{1}{|c|}{}&Time & \begin{tabular}{cc} Time & Speedup \end{tabular}  &\begin{tabular}{cc} Time & Speedup \end{tabular}&\begin{tabular}{cc} Time & Speedup \end{tabular}&\begin{tabular}{cc} Time & Speedup \end{tabular}&\begin{tabular}{cc} Time & Speedup \end{tabular}&\begin{tabular}{cc} Time & Speedup \end{tabular}&\begin{tabular}{cc} Time & Speedup \end{tabular}\\
\hline
\multicolumn{1}{|c|}{$8\times 10^{-3} $}&5.67e+00&\begin{tabular}{cc}  5.16e-01&11.0 \end{tabular}&\begin{tabular}{cc}4.52e+00 & 1.3 \end{tabular}&\begin{tabular}{cc}  9.78e-02&5.8e+01 \end{tabular}  &\begin{tabular}{cc}6.49e+01 &0.08 \end{tabular} &\begin{tabular}{cc} 7.17e+01&0.08 \end{tabular} & \begin{tabular}{cc} 5.28e+00 &1.1 \end{tabular} & \begin{tabular}{cc}  3.05e+00 &1.9e+00\end{tabular}\\
\multicolumn{1}{|c|}{$6\times 10^{-3} $}&7.69e+00&\begin{tabular}{cc}  1.33e+00&5.8 \end{tabular}&\begin{tabular}{cc}5.25e+00 & 1.5 \end{tabular}&\begin{tabular}{cc} 9.81e-02&7.8e+01 \end{tabular} &\begin{tabular}{cc} 9.49e+01&0.08 \end{tabular} &\begin{tabular}{cc} 8.84e+01&0.09 \end{tabular} & \begin{tabular}{cc}6.01e+00  &1.3 \end{tabular} & \begin{tabular}{cc} 3.05e+00 &2.5e+00 \end{tabular}\\
\multicolumn{1}{|c|}{$4\times 10^{-3} $}&1.30e+02&\begin{tabular}{cc}  1.67e+00&78.1 \end{tabular}&\begin{tabular}{cc}2.32e+01 & 5.6 \end{tabular}&\begin{tabular}{cc} 1.23e-01 &1.1e+03 \end{tabular} &\begin{tabular}{cc} 2.16e+02&0.6 \end{tabular} &\begin{tabular}{cc} 1.14e+02&1.1 \end{tabular} & \begin{tabular}{cc} 2.62e+01 & 5.0 \end{tabular} &\begin{tabular}{cc}3.08e+00  &4.2e+01 \end{tabular}\\
\multicolumn{1}{|c|}{$2\times 10^{-3} $}&4.83e+02&\begin{tabular}{cc}  4.07e+01&11.9 \end{tabular}&\begin{tabular}{cc}4.62e+01 & 10.5 \end{tabular}&\begin{tabular}{cc}  2.91e-01&1.7e+03 \end{tabular} &\begin{tabular}{cc} 6.56e+02&0.7 \end{tabular} &\begin{tabular}{cc} 6.16e+02&0.8\end{tabular} & \begin{tabular}{cc} 4.92e+01 &9.8 \end{tabular} &\begin{tabular}{cc} 1.41e+01 &3.4e+01 \end{tabular}\\
\multicolumn{1}{|c|}{$10^{-3} $}&9.22e+03&\begin{tabular}{cc}  1.80e+02&51.3 \end{tabular}&\begin{tabular}{cr}2.47e+02 & 37.3 \end{tabular}&\begin{tabular}{cc}  6.77e-01&1.4e+04 \end{tabular} &\begin{tabular}{cc} 2.92e+03&3.2 \end{tabular} &\begin{tabular}{cc} 2.46e+03&3.8 \end{tabular} & \begin{tabular}{cc} 2.61e+02 & 35.3 \end{tabular} & \begin{tabular}{cc} 1.45e+01 &6.4e+02 \end{tabular}\\
\multicolumn{1}{|c|}{$8\times 10^{-4} $}&1.50e+04&\begin{tabular}{cc}  2.60e+02&57.8 \end{tabular}&\begin{tabular}{cc}2.48e+02 & 60.5 \end{tabular}&\begin{tabular}{cc}  8.36e-01&1.8e+04 \end{tabular} &\begin{tabular}{cc} 4.63e+03&3.3 \end{tabular} &\begin{tabular}{cc} 3.53e+03&4.3 \end{tabular} & \begin{tabular}{cc} 2.62e+02 &57.3 \end{tabular} &\begin{tabular}{cc} 1.46e+01 &1.0e+03 \end{tabular}\\
\multicolumn{1}{|c|}{$6\times 10^{-4} $}&2.48e+04&\begin{tabular}{cc}  1.93e+03&12.9 \end{tabular}&\begin{tabular}{cc}4.13e+02 & 60.0 \end{tabular}&\begin{tabular}{cc}  1.97e+00&1.3e+04 \end{tabular} &\begin{tabular}{cc} 8.26e+03&3.0 \end{tabular} &\begin{tabular}{cc} 6.63e+03&3.8 \end{tabular} & \begin{tabular}{cc} 4.27e+02 &58.1 \end{tabular} &\begin{tabular}{cc} 7.76e+01 &3.2e+02 \end{tabular}\\
\multicolumn{1}{|c|}{$4\times 10^{-4} $}&5.68e+04&\begin{tabular}{cc}  4.12e+03&13.8 \end{tabular}&\begin{tabular}{cc}9.29e+02 & 61.1 \end{tabular} &\begin{tabular}{cc}  3.33e+00&1.7e+04 \end{tabular} &\begin{tabular}{cc} 1.69e+04&3.4 \end{tabular} &\begin{tabular}{cc} 1.62e+04&3.5 \end{tabular} & \begin{tabular}{cc} 9.43e+02 &60.2 \end{tabular} &\begin{tabular}{cc} 7.90e+01 &7.2e+02 \end{tabular}\\
\multicolumn{1}{|c|}{$2\times 10^{-4} $}&9.62e+05$^{\ast}$\!\!\!&\begin{tabular}{cc}  1.76e+04&54.6 \end{tabular}&\begin{tabular}{cc} 4.21e+03 & 228.5 \end{tabular} &\begin{tabular}{cc}  1.43e+01&6.7e+04 \end{tabular}&\begin{tabular}{cc} 7.13e+04&13.5 \end{tabular} &\begin{tabular}{cc} 6.38e+04&15.1 \end{tabular} & \begin{tabular}{cc}4.29e+03  & 224.2 \end{tabular} & \begin{tabular}{cc} 8.99e+01 &1.1e+04 \end{tabular}\\
\hline
\end{tabular}
}
\caption{CPU times in seconds together with speedups for the multilevel methods, for a variety of choices of $\epsilon$. For the interpolation post processing a common fine grid of level $\ell=5$ was used. The computational cost associated with a tolerance of $\epsilon = 2\times 10^{-4}$ for Monte Carlo was prohibitive; the entry in the table for this tolerance (with an asterisk) is an estimate.
}
\label{Tab:CPU_time_2}
\end{table}

Finally, in Table \ref{Tab:Sample_Size_MC}, we summarize the sample size estimations for both Monte Carlo and multilevel Monte Carlo sampling, considering both direct solve and surrogate methods, as quantified by \eqref{eq:MC_SampleSize} and \eqref{eq:MLMC_SampleSize}. The finest spatial mesh sizes used for direct solution and surrogate evaluation may not be identical. Surrogate-based approaches (both MC-FE and MLMC-FE) generally require a more stringent discretization error because they use a smaller splitting ratio ($\sqrt{0.5}/2$) in the nMSE error splitting compared to the larger splitting ratio of $\sqrt{0.5}$ used by the direct solver. Consequently, surrogate-based sampling typically requires a finer or at least the same finest spatial grid level as the direct solve. For example, in MC-FE sampling with tolerances $\epsilon=8\times 10^{-3}$, $6\times 10^{-3}$, $2\times 10^{-3}$, $6\times 10^{-4}$, $4\times 10^{-4}$, the direct solve method uses a mesh with one lower spatial grid level than the surrogate approach. Furthermore, both direct solver-based and surrogate-based sampling exhibit similar trends in sample size estimation for both Monte Carlo and multilevel Monte Carlo sampling. However, surrogate-based MLMC-FE sampling generally requires a slightly larger sample size. This is because, as shown in the right plot of Figure \ref{fig:CollectionPlot}, $V_\ell$ for the surrogate decreases slowly compared to the direct solver, coupled with the fact $W_\ell$ and $W_\ell^e$ increase at roughly similar rates, approximately 1.

\begin{remark}
As observed above, we found the cost of sampling by the direct method to be approximately $\frac{1.63\times 10^{-5}}{7.7\times 10^{-7}}\approx 21$ times greater if samples are obtained on grids of the same size $M_{\ell}$. Thus, if $C_d$ is the cost of computing a single direct solution, then the costs of computing $k$ samples are approximately

\vspace{.05in}

\begin{tabular}{ll}
$k C_d$   &\hspace{.24in} for sampling by direct solves, \\
$(25+k/21)C_d$   &\hspace{.25in} 
for sampling by surrogate,
\end{tabular}

\vspace{.05in}
\noindent
where the term ``$25C_d$'' reflects the overhead of the offline costs for 
the level-1 collocation sparse grid. This means that the surrogate is less expensive if $k>(21\cdot 25)/20 = 26.25$. It is clear from Table \ref{Tab:Sample_Size_MC} that typically many more samples are needed in practice. This comparison will also be affected by many factors: the direct solve entails solution of non-linear systems, which will be affected by the number of iterations, and as shown in Table \ref{Tab:Sample_Size_MC}, the grid sizes and number of samples will not in general be identical for the different strategies. Nevertheless, this simplified analysis clearly indicates the advantages of the surrogate for a simulation, despite its offline overhead.
\end{remark}

\begin{table}[ht]
\centering
\scalebox{0.675}{
\begin{tabular}{c|c|c|c|c|c|c||c|c|c|c|c|c||c|c|c|c|c|c||c|c|c|c|c|c|c|c|c|c|c|c|c|c|c|c|c|c|c|c|c|c|c|c|}
\cline{2-25}	
&\multicolumn{6}{|c||}{MC-FE + Direct solver} &\multicolumn{6}{|c||}{MC-FE + Surrogate}&\multicolumn{6}{|c||}{MLMC-FE + Direct solver}&\multicolumn{6}{|c|}{MLMC-FE + Surrogate} \\
\cline{2-25}	
&\multicolumn{6}{|c||}{ Level $\ell$} & \multicolumn{6}{|c||}{ Level $\ell$} & \multicolumn{6}{|c||}{ Level $\ell$} & \multicolumn{6}{|c|}{ Level $\ell$} \\
\hline
\multicolumn{1}{|c|}{\Large$\epsilon$}&0&1&2&3&4&5 &0&1&2&3&4&5 &0&1&2&3&4&5 &0&1&2&3&4&5\\
\hline
\multicolumn{1}{|c|}{$8\times 10^{-3} $}&&&5&&&& &&&4&& &10     &2     &2&&& &10     &2     &2     &2&& \\
\multicolumn{1}{|c|}{$6\times 10^{-3} $}&&&7&&&& &&&11&& &12     &3     &2&&& &10     &3     &2     &2&&\\
\multicolumn{1}{|c|}{$4\times 10^{-3} $}&&&&22&&& &&&14&& &32     &5     &2     &2&& &37     &8     &2     &2&&\\
\multicolumn{1}{|c|}{$2\times 10^{-3} $}&&&&83&&& &&&&83& &152    &26     &4     &2&& &151    &22     &4     &2     &2& \\
\multicolumn{1}{|c|}{$10^{-3} $}&&&&&322&& &&&&347& &691   &109    &18     &4     &2& &673   &134    &22     &5     &2&\\
\multicolumn{1}{|c|}{$8\times 10^{-4} $}&&&&&527&& &&&&501& &841   &129    &23     &3     &2&  &1017         &173          &29           &8           &2&\\
\multicolumn{1}{|c|}{$6\times 10^{-4} $}&&&&&869&& &&&&&901 &1610         &231          &40           &8           &2& &1867         &315          &51          &15           &4           &2\\
\multicolumn{1}{|c|}{$4\times 10^{-4} $}&&&&&1980&& &&&&&1995 &3791         &589         &104          &15           &3& &4160         &686         &108          &27           &6           &2\\
\multicolumn{1}{|c|}{$2\times 10^{-4} $}&&&&&& 8000$^{\ast}$\!\!& &&&&&8326 &15859        &2344         &375          &62          &13           &2 &17646        &2919         &435         &102          &20           &7\\
\hline
\end{tabular}
}
\caption{Optimal sample size estimation. In the case of Monte Carlo estimation on the finest grid, the computational cost associated with a tolerance of $\epsilon = 2\times 10^{-4}$ was prohibitive; the entry in the table for this tolerance (with an asterisk) is an estimate.}  
\label{Tab:Sample_Size_MC}
\end{table}

%
\subsection{Properties of geometric parameters}
%
We now look at some geometric quantities derived from the approximated \eqref{eq:QoI} that reflect the performance and accuracy of the surrogate and the overall effectiveness of the sampling methods. \\

\begin{figure}[ht!]\centering
\scalebox{.935}{\begin{tabular}{ccccc}
\multirow{ 4}{*}{\includegraphics[width=0.19\linewidth]{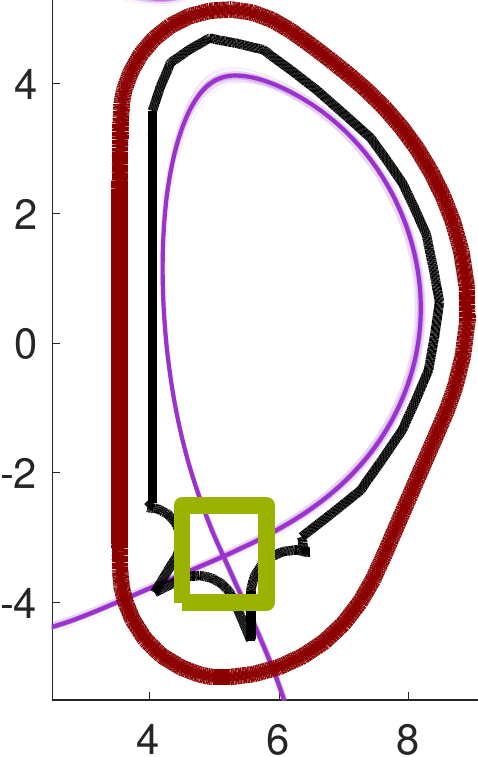} } & \includegraphics[width=0.19\linewidth]{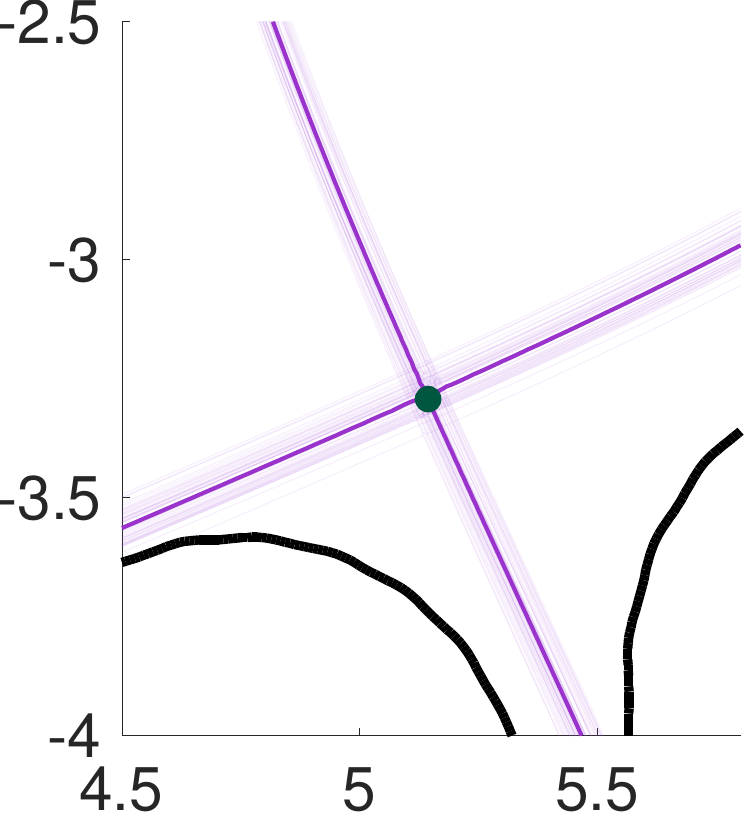} 
&\includegraphics[width=0.19\linewidth]{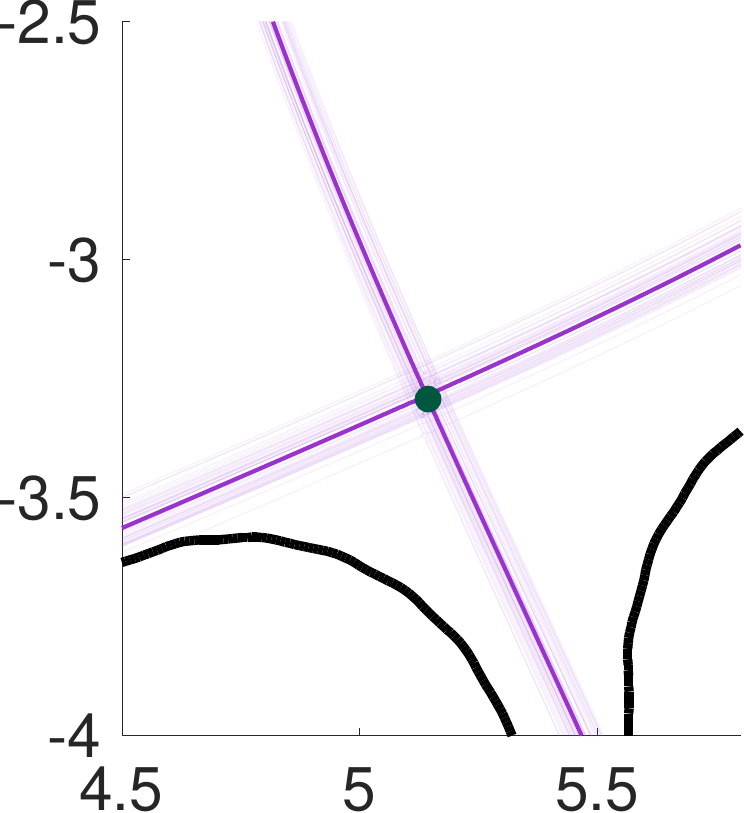}
&\includegraphics[width=0.19\linewidth]{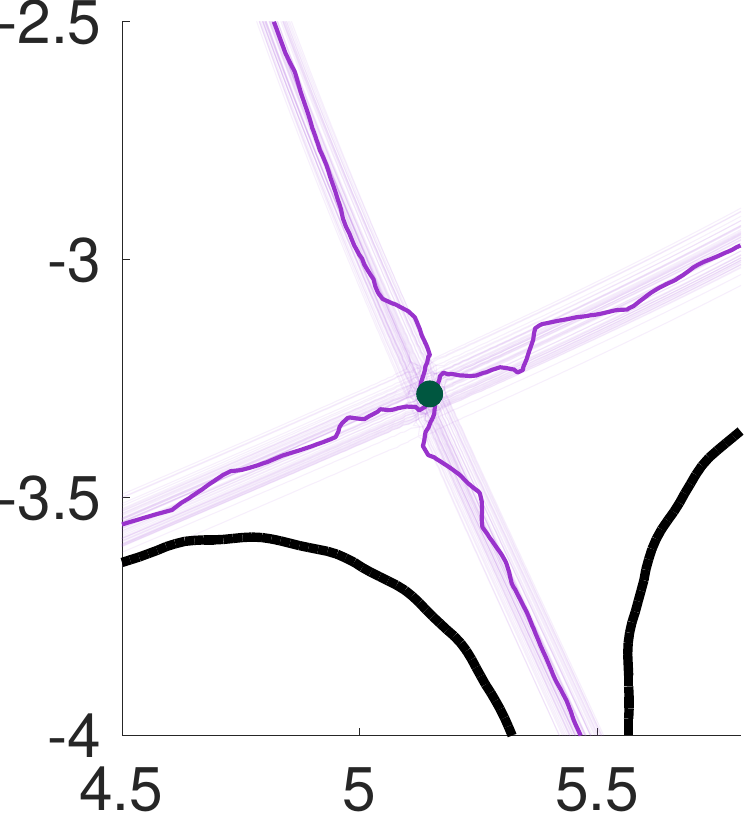} 
&\includegraphics[width=0.19\linewidth]{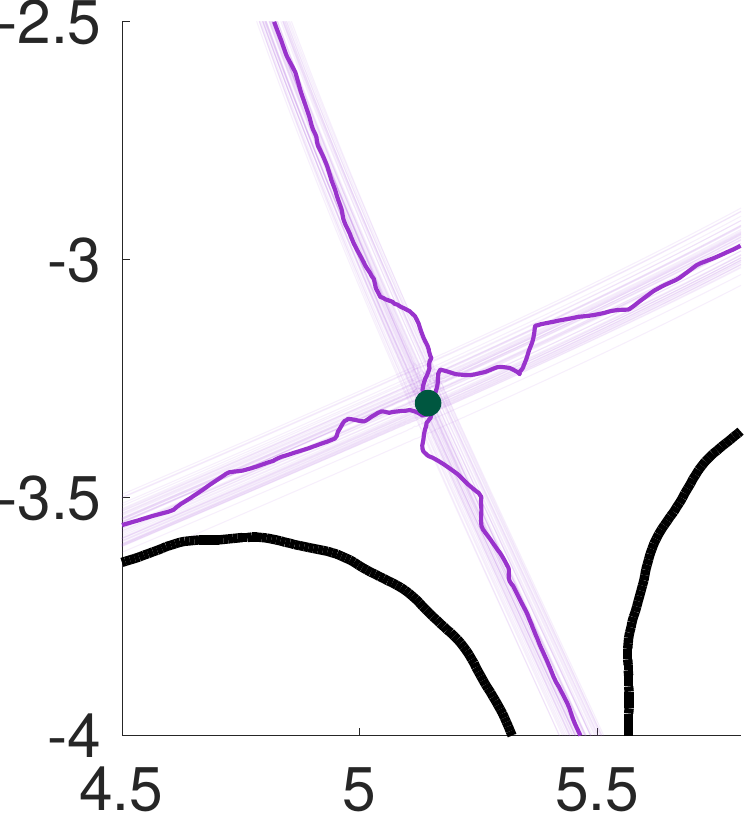}\\ 
& MC-FE& MC-FE &MLMC-FE  &MLMC-FE  \\
& Direct Solver & Surrogate  & Direct Solver & Surrogate  \\
& \includegraphics[width=0.19\linewidth]{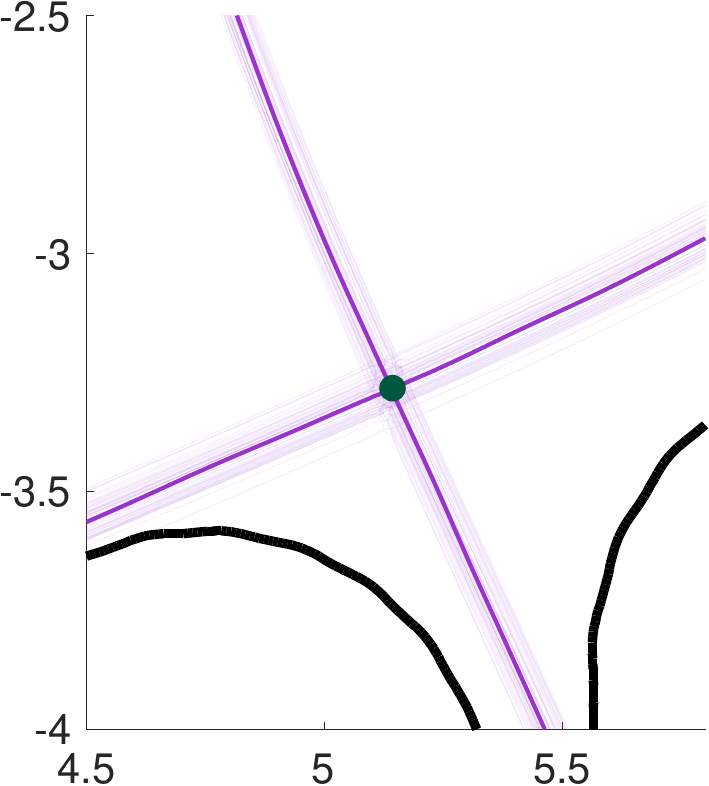} 
&\includegraphics[width=0.19\linewidth]{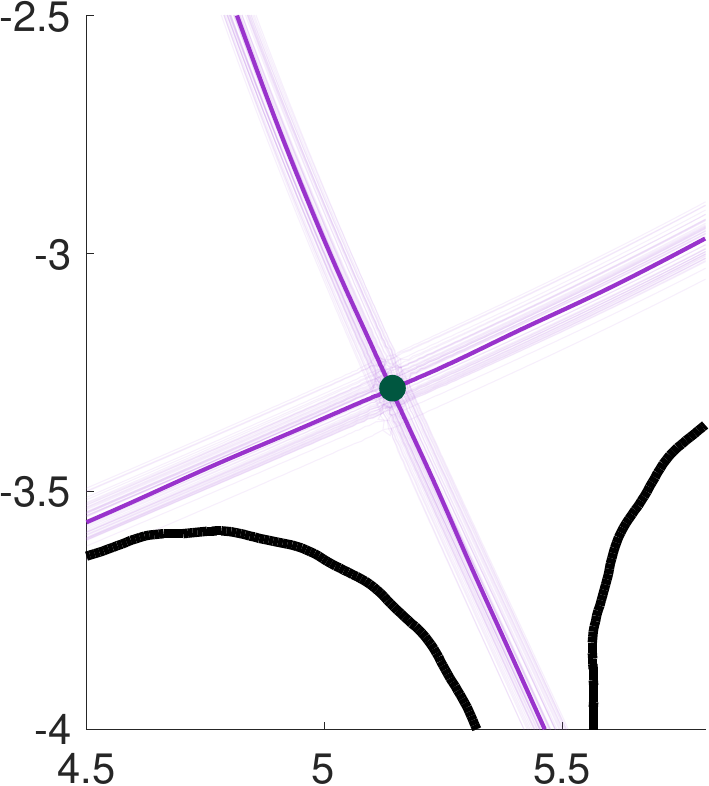}
&\includegraphics[width=0.19\linewidth]{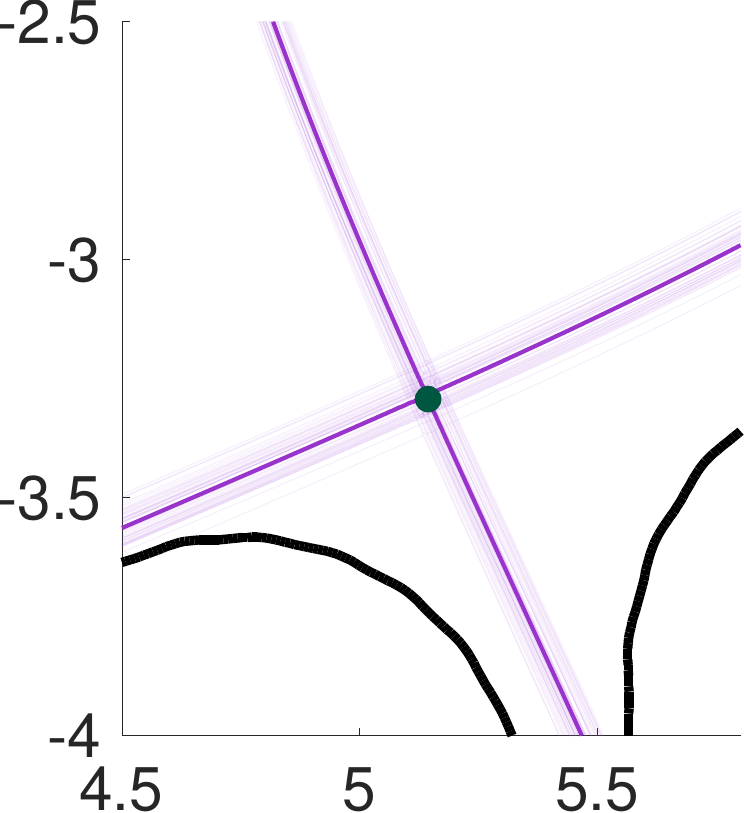}
&\includegraphics[width=0.19\linewidth]{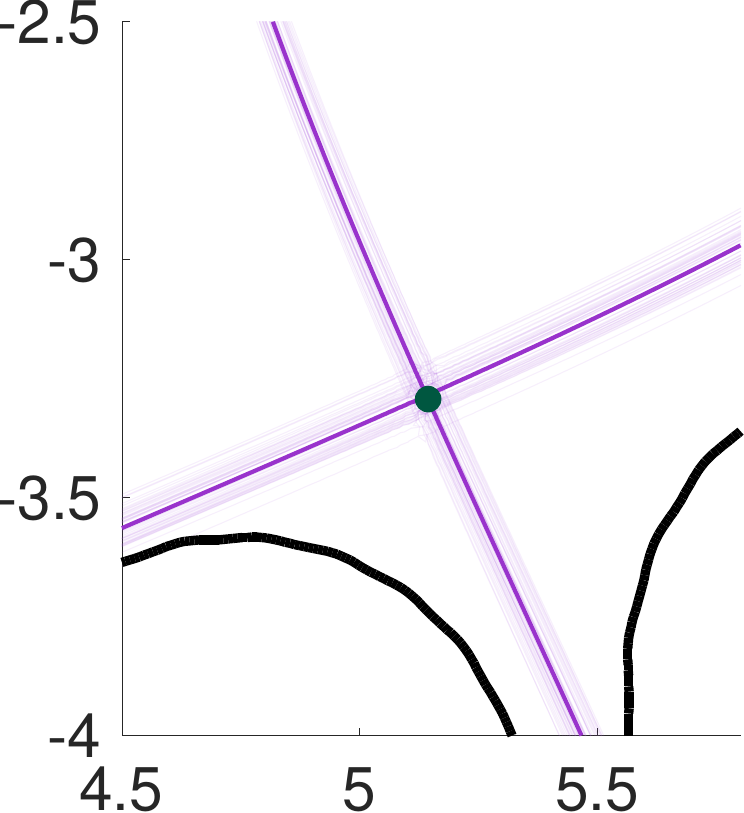}
\\[1ex]
Reactor & MLMC-FE& MLMC-FE &MLMC-FE  &MLMC-FE  \\
cross section & Direct Solver & Surrogate  & Direct Solver & Surrogate  \\
 & Heat flow & Heat flow & Interpolation & Interpolation\\[-0.5ex]
\end{tabular}}
\caption{The overlayed plasma boundaries of 50 random realizations are displayed as violet curves. The solid violet line is the plasma boundary of the expected poloidal flux generated with tolerance $\epsilon=4\times 10^{-4}$. On the leftmost panel, the inner and outer walls of the reactor are displayed in solid black and dark red respectively, while the green square highlights the zoomed--in in the remaining images. The dark green dots are the x-points of the expected solution. All simulations were performed using the discretization level $\ell=5$ on geometry-conforming uniform meshes.}
\label{fig:QoI_plot}
\end{figure}
\vspace{.05in}

\noindent \textbf{Plasma boundary.} 
Figure \ref{fig:QoI_plot} shows the plasma boundary of the expected poloidal flux in \eqref{eq:QoI} for both direct solve and surrogate evaluation. As discussed in \cite{ElLiSa:2023}, Monte Carlo sampling with direct computation on a single spatial grid results in a smooth plasma boundary, and the use of a surrogate achieves the same outcome, as seen in the second and third columns of the top row in the figure. However, multilevel Monte Carlo sampling introduces distortions in the profile of the plasma boundary due to the interaction of samples defined across different grids during the accumulation of sample corrections. This is evident in the last two columns of the top row in Figure \ref{fig:QoI_plot}. Thus, the straightforward use of multilevel methods in this setting exhibits a trade-off between dramatically increased efficiency and somewhat decreased accuracy.\\

\noindent\textbf{Post--processing strategies.}
We propose two strategies to eliminate this distortion. The principled approach is to interpolate the sample corrections to a common fine grid (near the x-point, solutions computed this way look exactly like those in the bottom-right two images of Figure \ref{fig:QoI_plot}), but as shown in Table \ref{Tab:CPU_time_2}, this comes at significant expense. A second strategy to filter out the spurious high--frequency components of the multi--level approximation is to take advantage of the smoothing properties of the heat operator by subjecting the approximation to a simple heat flow. This approach avoids the expense of interpolating every realization into a common grid.

More precisely, let us denote the multi--level approximation to the expectation as $\psi$ and by
\[
\mathbf M := \int_\Omega \phi_i\phi_j \,dr\,dz \qquad \text{ and } \qquad \mathbf S := \int_\Omega \nabla\phi_i\cdot\nabla\phi_j \,dr\,dz
\]
the mass and stiffness matrices for the nodal basis functions $\{\phi_i\}$ of the mesh where the approximated expectation $\psi$ was computed. Then, the filtered approximation $\psi^*$ is given by
\[
\psi^* = \left(\left(\mathbf M +\kappa\alpha \mathbf S\right)^{-1}\mathbf M\right)^n\psi\,.
\]
The matrix $\mathbf A:=\left(\mathbf M +\kappa\alpha \mathbf S\right)$ can be precomputed once and then the system $\mathbf A^{-1}\mathbf{M}\psi$ solved $n$ times to obtain the filtered expectation $\psi^*$. The post--processing formula above amounts to an implicit--in--time Galerkin discretization of the initial boundary value problem
\[
\partial_t\psi^* = \alpha\Delta\psi^* \;\; \text{ in } \Omega\times (0,n\kappa]\,, \qquad
\partial_\nu\psi^* = 0 \;\; \text{ on } \Gamma \times (0,n\kappa]\,,\qquad
\psi^* = \psi \;\; \text{ on } \overline\Omega \times \{0\}\,,
\]
where $0<\kappa$ is the size of the pseudo time step, $0<\alpha$ is the pseudo diffusivity, and $n$ is the number of pseudo time steps. In our experiments we used $\kappa = 2 \times 10^{-2}$, $\alpha=0.1 $, and $n=5$.  

As depicted in the second and third columns in the bottom row of Figure \ref{fig:QoI_plot}, after the post--processing step, the distortion of the boundary essentially disappears at a computational cost that is significantly smaller than that of interpolating the samples into a common grid. As shown in Table \ref{Tab:CPU_time_2}, the associated speed--ups (final column) are not quite as large as those of the un--processed strategy (seventh column) but still significant compared to those of the multi--level direct solver (fourth column). This is specially true for the more challenging cases of stringent tolerances, where the computational time was decreased by factors ranging between $10^2$ and $10^4$. As we will discuss in the next subsection, the post--processing step has no detrimental effect on the determination of the relevant geometric parameters of the plasma.\\

\noindent \textbf{Geometric descriptors.}
Table \ref{Tab: QoI_GeoInfo} summarizes some values of geometric parameters derived from the plasma boundaries of the expected plasma field for various methods.  In the table, we set the outcome of Monte Carlo sampling with direct computations as the benchmark. We find that the surrogate--enhanced Monte Carlo produces parameters identical to the direct--solve results up to two decimal places. In contrast, MLMC sampling with both direct computation and surrogate methods on geometry--conforming grids aligns only up to one decimal place, due to plasma boundary distortions from extrapolation errors on non--nested grids. However, when using a common grid for interpolation, the geometric descriptors for MLMC sampling, both direct solve and surrogate, incur a relative error significantly smaller than 1\%. A similarly accurate behavior is observed with both filtering strategies for the multi--level approximations.

\begin{table}[ht]
	\centering
			\scalebox{0.72}{
		\begin{tabular}{c|c|c|c|c|c|c|c|c|c|c|c|c|}
			\cline{2-9}
                &\multicolumn{1}{c|}{}& & & & Interpolation & Interpolation & Heat flow & Heat flow \\
				&\multicolumn{1}{c|}{MC-FE}&MC-FE&MLMC-FE&MLMC-FE &MLMC-FE &MLMC-FE &MLMC-FE&MLMC-FE \\
				&\multicolumn{1}{c|}{Direct Solver}&Surrogate& Direct solver &  Surrogate & Direct solver & Surrogate & Direct solver&  Surrogate\\
			\hline
			\multicolumn{1}{|c|}{x point}&(5.14,-3.29)&(5.14,-3.29)&(5.14,-3.29)&(5.14,   -3.30)&(5.14,-3.29)&(5.14,-3.29) &(5.14,-3.28) &(5.14,-3.28)\\
			\hline
			\multicolumn{1}{|c|}{magnetic axis}&(6.41,0.61)&(6.41,0.61)&(6.44,0.56)&(6.44,    0.56)&(6.41,0.61)&(6.41,0.61) &(6.41,0.60)  &(6.41,0.60) \\
			\hline
			\multicolumn{1}{|c|}{strike} &(4.16,-3.71)&(4.16,-3.71)&(4.16,-3.71)&(4.16,   -3.71)&(4.16,-3.71)&(4.16,-3.71) &(4.16,-3.71) &(4.16,-3.71)\\
			\multicolumn{1}{|c|}{points}&(5.56,-4.22)&(5.56,-4.22)&(5.56,-4.22)&(5.56,   -4.22)&(5.56,-4.22)&(5.56,-4.22) &(5.56,-4.22) &(5.56,-4.22)\\
			\hline
			\multicolumn{1}{|c|}{inverse aspect ratio} &0.32&0.32&0.32&0.32&0.32&0.32 &0.32 &0.32\\
			\hline
			\multicolumn{1}{|c|}{elongation} &1.86&1.86&1.87&1.87&1.86&1.86 &1.86 &1.86\\
			\hline
			\multicolumn{1}{|c|}{upper triangularity}&0.43&0.43&0.43&0.43&0.43&0.43 &0.44 &0.44\\
			\hline
			\multicolumn{1}{|c|}{lower triangularity} &0.53&0.53&0.53&0.53&0.53&0.53 &0.53 &0.53\\
			\hline
	\end{tabular}
  }
	\caption{Geometric parameters of the expected poloidal flux $u$. The results are generated with a target normalized mean squared error of $\epsilon=4\times 10^{-4}$. In the case of the interpolation post--processing strategy, the results were interpolated into a common fine grid of level $L=5$.} 
	\label{Tab: QoI_GeoInfo}
\end{table}

 \section{Concluding remarks}  \label{sec:Conclusion}
This paper proposes a surrogate-based MLMC-FE sampling strategy to estimate the expectation of the magnetic field in the Grad-Shafranov free boundary problem with high-dimensional uncertainties in current intensities. Cost analyses demonstrate the effects of using surrogate approximations defined by sparse-grid collocation methods to the solutions of the nonlinear systems of equations arising in the model, or of using multilevel Monte Carlo methods to reduce the cost of direct solves,  and of combining these two ideas to use sparse-grid surrogates together with multilevel methods. Computational experiments with the Grad-Shafranov equation demonstrate that each of these approaches leads to reductions in costs to perform simulations, with very dramatic cost reductions (approaching factors of $10^4$) obtained from the combined methods. These savings come with some sacrifice of accuracy for the surrogate solutions, most notably for multilevel spatial methods, although the quantitative values of important measures in the model agree to two digits. Two filtering strategies are offered to counteract the loss of accuracy in the multi--level estimations. The heat flow smoothing is by far the the most efficient, as it offers speed--ups ranging from $10^{2}$ to $10^4$ and produces estimates that fall within less of 1\% of those of a direct Monte Carlo estimation on the finest spatial grid.
%
\section{Acknowledgements}
Jiaxing Liang was partially supported by the U. S. Air Force Research Laboratory through the grant AFOSR FA9550-22-1-0004. Tonatiuh S\'anchez-Vizuet was partially supported by the U. S. National Science Foundation through the grant  NSF-DMS-2137305.\\

\noindent\textbf{Statement of code availability}\\
The code used for the numerical experiments in this article has three components:
\begin{enumerate}
\item The free--boundary solver used to obtain the ``truth" solutions for the construction of the surrogate, {\tt FEEQS.m} \cite{Heumann:feeqsm} is a Matlab implementation of the code {\tt CEDRES++} \cite{CEDRES}, kindly provided by Holger Heumann. The authors of this communication are not at liberty of sharing the source. However, the interested reader is referred to the contact information found at \cite{Heumann:feeqsm, CEDRES} for code requests.
\item The code producing the sparse grid surrogates {\tt SPINTERP}, due to Klimke and Wohlmuth, is available from the Association of Computing Machinery's Transactions on Mathematical Software as the supplementary material for the article \cite{KlBa:2005}.
\item The routines used to interface between {\tt SPINTERP} and {\tt FEEQS.m}, the implementation of Algorithm \ref{algo:MLMC_Algo_CorrectionVersion}, and  the post--processing tools used to extract geometric information, perform heat flow regularization, statistical analysis, visualization, etc. are due to the authors of this communication and can be made available upon request.
\end{enumerate}
%

\bibliographystyle{abbrv}
\bibliography{references}

\begin{thebibliography}{10}

\bibitem{BaNoTe:2007}
I.~Babu\v{s}ka, F.~Nobile, and R.~Tempone.
\newblock A stochastic collocation method for elliptic partial differential
  equations with random input data.
\newblock {\em SIAM Journal on Numerical Analysis}, 45(3):1005--1034, 2007.

\bibitem{BaScZo:2011}
A.~Barth, C.~Schwab, and N.~Zollinger.
\newblock Multi-level {M}onte {C}arlo finite element method for elliptic {PDE}s
  with stochastic coefficients.
\newblock {\em Numerische Mathematik}, 119(1):123--161, 2011.

\bibitem{BaNoRi:2000}
V.~Barthelmann, E.~Novak, and K.~Ritter.
\newblock High dimensional polynomial interpolation on sparse grids.
\newblock {\em Advances in Computational Mathematics}, 12:273--288, 2000.

\bibitem{Braess2007}
D.~Braess.
\newblock {\em Finite Elements: Theory, Fast Solvers, and Applications in Solid
  Mechanics}.
\newblock Cambridge University Press, 3rd. edition, Apr. 2007.

\bibitem{Carnicer1990}
J.~Carnicer and M.~Gasca.
\newblock Evaluation of multivariate polynomials and their derivatives.
\newblock {\em Mathematics of Computation}, 54(189):231–243, 1990.

\bibitem{Clenshaw1955}
C.~W. Clenshaw.
\newblock A note on the summation of {C}hebyshev series.
\newblock {\em Mathematics of Computation}, 9(51):118–120, 1955.

\bibitem{ClCu:1960}
C.~W. Clenshaw and A.~R. Curtis.
\newblock A method for numerical integration on an automatic computer.
\newblock {\em Numer. Math.}, 2:197--205, 1960.

\bibitem{ClGiScTe:2011}
K.~A. Cliffe, M.~B. Giles, R.~Scheichl, and A.~L. Teckentrup.
\newblock Multilevel {M}onte {C}arlo methods and applications to elliptic
  {PDE}s with random coefficients.
\newblock {\em Computing and Visualization in Science}, 14(1):3--15, 2011.

\bibitem{Da:2006}
T.~A. Davis.
\newblock {\em Direct Methods for Sparse Linear Systems}, volume~2 of {\em
  Fundamentals of Algorithms}.
\newblock Society for Industrial and Applied Mathematics (SIAM), Philadelphia,
  PA, 2006.

\bibitem{DePe2003}
J.~Delgado and J.~Peña.
\newblock A linear complexity algorithm for the {B}ernstein basis.
\newblock In {\em 2003 International Conference on Geometric Modeling and
  Graphics, 2003. Proceedings}, pages 162--167, 2003.

\bibitem{DePe2007}
J.~Delgado and J.~M. Peña.
\newblock A corner cutting algorithm for evaluating rational {B}ézier surfaces
  and the optimal stability of the basis.
\newblock {\em SIAM Journal on Scientific Computing}, 29(4):1668–1682, Jan.
  2007.

\bibitem{EiMeNe:2016}
M.~Eigel, C.~Merdon, and J.~Neumann.
\newblock An adaptive multilevel {M}onte {C}arlo method with stochastic bounds
  for quantities of interest with uncertain data.
\newblock {\em SIAM/ASA Journal on Uncertainty Quantification},
  4(1):1219--1245, 2016.

\bibitem{ElLiSa:2022}
H.~C. Elman, J.~Liang, and T.~S\'{a}nchez-Vizuet.
\newblock Surrogate approximation of the {G}rad-{S}hafranov free boundary
  problem via stochastic collocation on sparse grids.
\newblock {\em Journal of Computational Physics}, 448:110699, 20, 2022.

\bibitem{ElLiSa:2023}
H.~C. Elman, J.~Liang, and T.~S\'{a}nchez-Vizuet.
\newblock Multilevel {M}onte {C}arlo methods for the {G}rad-{S}hafranov free
  boundary problem.
\newblock {\em Computer Physics Communications}, 298:109099, 2024.

\bibitem{FaHe:2017}
B.~Faugeras and H.~Heumann.
\newblock {FEM-BEM} coupling methods for {T}okamak plasma axisymmetric
  free-boundary equilibrium computations in unbounded domains.
\newblock {\em Journal of Computational Physics}, 343:201 -- 216, 2017.

\bibitem{Ge:1973}
A.~George.
\newblock Nested dissection of a regular finite element mesh.
\newblock {\em SIAM J. Numer. Anal.}, 10:345--363, 1973.

\bibitem{Gi:2008}
M.~B. Giles.
\newblock Multilevel {M}onte {C}arlo path simulation.
\newblock {\em Operations Research}, 56(3):607--617, 2008.

\bibitem{Gi:2015}
M.~B. Giles.
\newblock Multilevel {M}onte {C}arlo methods.
\newblock {\em Acta Numerica}, 24:259--328, 2015.

\bibitem{GoVaCh:2013}
G.~H. Golub and C.~F. Van~Loan.
\newblock {\em Matrix Computations}.
\newblock Johns Hopkins Studies in the Mathematical Sciences. Johns Hopkins
  University Press, Baltimore, MD, fourth edition, 2013.

\bibitem{GrRu:1958}
H.~Grad and H.~Rubin.
\newblock Hydromagnetic equilibria and force{-}free fields.
\newblock In {\em Proc. Second international conference on the peaceful uses of
  atomic energy, Geneva}, volume 31,190, New York, Oct 1958. United Nations.

\bibitem{Gr:1999}
V.~Grandgirard.
\newblock Modelisation de l'equilibre d'un plasma de tokamak.
\newblock Technical report, Universit\'e de {F}ranche-{C}omt\'e, 1999.

\bibitem{Heumann:feeqsm}
H.~Heumann.
\newblock {FEEQS.M}.
\newblock http://www-sop.inria.fr/members/Holger.Heumann/Software.html.

\bibitem{CEDRES}
H.~Heumann, J.~Blum, C.~Boulbe, B.~Faugeras, G.~Selig, J.-M. Ané, S.~Brémond,
  V.~Grandgirard, P.~Hertout, E.~Nardon, and et~al.
\newblock Quasi-static free-boundary equilibrium of toroidal plasma with
  {CEDRES++}: Computational methods and applications.
\newblock {\em Journal of Plasma Physics}, 81(3):905810301, 2015.

\bibitem{HoGi1819}
W.~G. Horner and D.~Gilbert.
\newblock {XXI}. {A} new method of solving numerical equations of all orders,
  by continuous approximation.
\newblock {\em Philosophical Transactions of the Royal Society of London},
  109:308--335, 1819.

\bibitem{KhPaHe:2020}
A.~Khodadadian, M.~Parvizi, and C.~Heitzinger.
\newblock An adaptive multilevel {M}onte {C}arlo algorithm for the stochastic
  drift-diffusion-{P}oisson system.
\newblock {\em Computer Methods in Applied Mechanics and Engineering},
  368:113163, 23, 2020.

\bibitem{KlBa:2004}
A.~Klimke, K.~Willner, and B.~Wohlmuth.
\newblock Uncertainty modeling using fuzzy arithmetic based on sparse grids:
  applications to dynamic systems.
\newblock {\em International Journal of Uncertainty, Fuzziness and
  Knowledge-Based Systems}, 12(6):745--759, 2004.

\bibitem{KlBa:2005}
A.~Klimke and B.~Wohlmuth.
\newblock Algorithm 847: spinterp: piecewise multilinear hierarchical sparse
  grid interpolation in {MATLAB}.
\newblock {\em Association for Computing Machinery. Transactions on
  Mathematical Software}, 31(4):561--579, 2005.

\bibitem{Li:2024}
J.~Liang.
\newblock {\em Efficient Computational Algorithms for Magnetic Equilibrium in a
  Fusion Reactor}.
\newblock Phd thesis, University of Maryland, 2024.

\bibitem{LuSc:1957}
R.~L\"{u}st and A.~Schl\"{u}ter.
\newblock Axialsymmetrische magnetohydrodynamische
  {G}leichgewichtskonfigurationen.
\newblock {\em Z. Naturf}, 12a:850--854, 1957.

\bibitem{LuBr:1982}
J.~Luxon and B.~Brown.
\newblock Magnetic analysis of non-circular cross-section tokamaks.
\newblock {\em Nuclear Fusion}, 22(6):813--821, jun 1982.

\bibitem{MaNi:2009}
X.~Ma and N.~Zabaras.
\newblock An adaptive hierarchical sparse grid collocation algorithm for the
  solution of stochastic differential equations.
\newblock {\em Journal of Computational Physics}, 228(8):3084--3113, 2009.

\bibitem{NoTeWe:2008}
F.~Nobile, R.~Tempone, and C.~G. Webster.
\newblock A sparse grid stochastic collocation method for partial differential
  equations with random input data.
\newblock {\em SIAM Journal on Numerical Analysis}, 46(5):2309--2345, 2008.

\bibitem{NoTe:2015}
F.~Nobile and F.~Tesei.
\newblock A multi level {M}onte {C}arlo method with control variate for
  elliptic {PDE}s with log-normal coefficients.
\newblock {\em Stochastic Partial Differential Equations. Analysis and
  Computations}, 3(3):398--444, 2015.

\bibitem{Shafranov:1958}
V.~D. Shafranov.
\newblock On magnetohydrodynamical equilibrium configurations.
\newblock {\em Soviet Physics JETP}, 6:545--554, 1958.

\bibitem{Sh:2002}
J.~R. Shewchuk.
\newblock Delaunay refinement algorithms for triangular mesh generation.
\newblock {\em Computational Geometry}, 22(1-3):21--74, May 2002.

\bibitem{Sm:1963}
S.~A. Smolyak.
\newblock Quadrature and interpolation formulae on tensor products of certain
  function classes.
\newblock {\em Doklady Akademii Nauk SSSR}, 148:1042--1045, 1963.

\bibitem{TeJaWe:2015}
A.~L. Teckentrup, P.~Jantsch, C.~G. Webster, and M.~Gunzburger.
\newblock A multilevel stochastic collocation method for partial differential
  equations with random input data.
\newblock {\em SIAM/ASA Journal on Uncertainty Quantification},
  3(1):1046--1074, 2015.

\bibitem{TeScGiUl:2013}
A.~L. Teckentrup, R.~Scheichl, M.~B. Giles, and E.~Ullmann.
\newblock Further analysis of multilevel {Monte Carlo} methods for elliptic
  {PDEs} with random coefficients.
\newblock {\em Numerische Mathematik}, 125(3):569--600, Mar. 2013.

\end{thebibliography}
\end{document}